\documentclass[journal]{IEEEtran} 

\usepackage{epsfig}
\usepackage{calc}

\usepackage{amsmath}
\usepackage{amssymb}
\usepackage{amsfonts}
\usepackage{epsfig}

\newcommand{\cnoi}{\mathbb{C}\mathcal{N}(0,1)}

\newcounter{propo}

\newtheorem{defn}{Definition}
\newtheorem{thm}{Theorem}
\newtheorem{prop}[propo]{Proposition}
\newtheorem{lem}[thm]{Lemma}

\newtheorem{note}{Remark}
\newtheorem{ex}{{\em Example}}

\newcommand{\epf}{\hfill $\Box$}

\newcommand{\beq}{\begin{equation}}
\newcommand{\eeq}{\end{equation}}

\newcommand{\bea}{\begin{eqnarray}}
\newcommand{\eea}{\end{eqnarray}}
\newcommand{\bean}{\begin{eqnarray*}}
\newcommand{\eean}{\end{eqnarray*}}

\newcommand{\bit}{\begin{itemize}}
\newcommand{\eit}{\end{itemize}}

\begin{document}
\title{Approximately universal optimality over several dynamic and non-dynamic cooperative diversity schemes for wireless networks}

\author{Petros Elia and P. Vijay Kumar
\thanks{Petros Elia and P. Vijay Kumar are with the Department of EE-Systems,
University of Southern California, Los Angeles, CA 90089 ({\tt
\{elia,vijayk\}@usc.edu}).  This work was carried out while P.
Vijay Kumar was on a leave of absence at the Indian Institute of
Science Bangalore. }}

\maketitle\thispagestyle{empty}
\bibliographystyle{ieeetran}
\begin{abstract}
In this work we explicitly provide the first ever optimal, with
respect to the Zheng-Tse diversity multiplexing gain (D-MG)
tradeoff, cooperative diversity schemes for wireless relay
networks.  The schemes are based on variants of perfect space-time
codes and are optimal for any number of users and all
statistically symmetric (and in some cases, asymmetric) fading
distributions.

We deduce that, with respect to the D-MG tradeoff, channel
knowledge at the intermediate relays and infinite delay are
unnecessary.  We also show that the non-dynamic selection decode
and forward strategy, the non-dynamic amplify and forward, the
non-dynamic receive and forward, the dynamic amplify and forward
and the dynamic receive and forward cooperative diversity
strategies allow for exactly the same D-MG optimal performance.
\end{abstract}

\section{Introduction \label{sec:RelayJournal_Introduction}}
The emerging need for reliable communications of large quantities
of data, at high rates, between small and independent users with
small power supplies, no antenna arrays and minimal computational
capabilities, brought to the fore cooperative-diversity in
wireless networks, where distributed users relay messages for one
another, in order to combat the fading and additive noise that
hinters their joint communication, hence improving the overall
quality of service. Cooperation can be achieved by relating each
user with a segment of a point-to-point communication scheme and
essentially having the intermediate relays manipulate the signal
from the information source in a way that the received signal at
the final destination relates to that of a point to point
transmission with multiple transmit antennas. Network outage
analysis provides for the fundamental limits of the network's
performance.

\subsection{Structure of the paper}
In Section \ref{sec:RelayJournal_Introduction} we will describe
the cooperative-diversity setup and the related notation that will
be used in the rest of the paper, will briefly introduce the
existing cooperative diversity strategies, state the existing
performance bounds and quickly go over the basic aspects of the
D-MG tradeoff.

In Section \ref{sec:RelayJournal_ND-SDAF}, Theorem
\ref{thm:optimalDMG of decode and forward}, we will present the
expression for optimal D-MG performance for the non-dynamic
selection-decode-and-forward strategy (ND-SDAF
\cite{LanemanII,LanWorTSEIEEE}) and will provide the exact coding
methodology that achieves this optimality.  In proving the
scheme's optimality, we will use the existence of sets of
approximately universal \cite{TavVisUniversal_2005} codes which
hold common elements, and which maintain approximate universality
even when their structure is altered.  We will also generalize
with respect to network topology and fading statistics.

In Section \ref{sec:RelayJournal_ND-RAF} we will present a variant
of the non-dynamic linear-processing relay network
\cite{JingHass04b}.  Unlike with ND-SDAF, this non-dynamic
receive-and-forward (ND-RAF) scheme requires for coding
distribution across the relays. We will then proceed to analyze
the second stage equivalent (`two-product') channel and prove its
partial D-MG equivalence to the Rayleigh fading channel. Using the
existence of sets of approximately universal codes with joint
elements, we will present in Theorem \ref{thm:optimalDMG of
ND-RAF} the optimal performance for the ND-AAF and ND-RAF scheme
and the exact coding methodology that achieves this performance.
As a guideline for other potential constructions, we will
introduce one by one all the necessary conditions for achieving
optimality. A practical scheme, more suited for a network with a
large number of users and which provides for a plethora of
practical advantages, will also presented. Finally, a base station
network setup will be presented together with some new variants of
perfect codes that have the potential to render cooperation
fruitful at the lowest possible SNR.

In Theorem \ref{thm:D AAF all relays} of Section
\ref{sec:RelayJournal_D-RAF}, we present D-MG optimal schemes for
the dynamic amplify-and-forward and the dynamic
receive-and-forward strategies (D-AAF, D-RAF
\cite{ElGamalDUplex}). Optimality holds for any number of users,
any set of channel statistics and any statistical asymmetry.

Section \ref{sec:RelayJournal_Comparison of schemes} offers a
comparison of the cooperative diversity schemes, and through
Theorem \ref{thm:all schemes are the same} it is concluded that
all the above strategies have high-SNR outage regions of the same
volume and thus achieve the same D-MG optimal performance.  We
point out that given statistical symmetry, the ND-RAF scheme
offers this same optimal performance at a reduced delay, reduced
decoding complexity, reduced signalling complexity, minimal
computational efficiency and the highest ease for network
deployment. Section \ref{sec:relay journal simulations} presents
some simulations. The rest of the sections are appendices.

We begin we a general description of the network.
\subsection{Describing the network} In \cite{LanemanII},
the authors describe the case where a set $$\mathcal{R} = \{
R_1,R_2,\cdots,R_{n},R_{n+1}\}$$ of $n+1$ different
terminals/relays, cooperate in their effort to communicate with
each other. Each relay has the ability to communicate over $n+1$
different orthogonal frequencies $\mathcal{F} = \{
\nu_1,\nu_2,\cdots,\nu_{n},\nu_{n+1}\}$. A certain relay $R_i$,
wanting to communicate with relay $d(R_i)$, broadcasts its
original information over frequency $\nu_i$.  Depending on the
availability of each intermediate relay, the set
\begin{equation}\label{eq:set of cooperating relays}D(R_i)\subset
\{ \mathcal{R} \setminus \{R_i \cup
 d(R_i) \}\}\end{equation} is then the set of all intermediate relays
that cooperate with $R_i$.  Consequently, each relay $R_j\in
D(R_i)$ transmits a possibly modified version of the received
signal over frequency $\nu_i$. By the end of the transmission,
$d(R_i)$ has received the information from $R_i$ over frequency
$\nu_i$, in a form of a superposition of faded versions of signals
originating from $R_i$ and from $D(R_i)$.

We will focus on the case where communication takes place in the
presence of additive receiver noise, and in the presence of
spatially independent quasi-static fading.  Furthermore, we will
assume complete knowledge of the fading channel at the receiver of
the final destination, and depending on the cooperative diversity
strategy, we will assume complete knowledge or absolutely no
knowledge of the channel at the receivers of the intermediate
relays.  Finally, the half-duplex condition is imposed, due to
practical considerations such as the large ratio between the
transmission and reception powers at the relay antennas
(\cite{LanemanII,LanWorTSEIEEE,ElGamalDUplex}).
\paragraph{Instance of a network}
From \cite{LanemanII} we see that without loss of generality we
can analyze the overall network performance just be focusing on a
snapshot of the network, as shown in Figure \ref{fig:General
snapshot}, where $S$ is now the information source, $D$ the
destination, $R_2,\cdots,R_{n}$ are the intermediate relays,
\begin{figure}[h]
\begin{center}
\begin{center}\includegraphics[width=0.9\columnwidth]{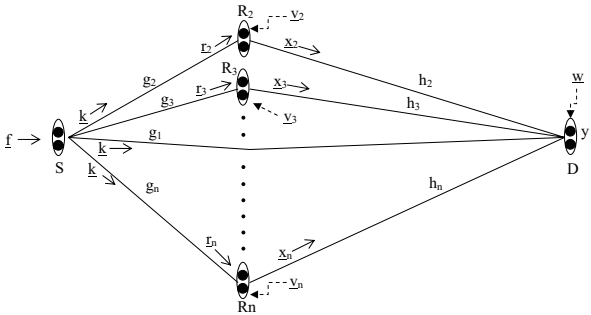}\end{center}
\caption{Snapshot of wireless network, where terminal $S$ utilizes
its peers ($R_2,\cdots,R_{n}$) for communicating with $D$.
\label{fig:General snapshot}}
\end{center}
\end{figure}
$g_i$ is the fading coefficient between $S$ and intermediate relay
$R_i$, $h_i$ is the fading from $R_i$ to $D$ and $h_1$ is the
fading coefficient from $S$ to $D$. We consider $h_i,g_i$ to be
independently distributed circularly symmetric $\cnoi$ random
variables, remaining constant throughout the transmission. Vectors
$\underline{v}_i$ and $\underline{w}$ contain the elements
$v_{i,j}$ and $w_{j}$ corresponding to the additive receiver noise
respectively affecting $R_i$ and $D$ at time $t=j$. Unless we
state otherwise, we ask that all $v_{i,j}$ and $w_{j}$ be
independently distributed $\cnoi$ random variables. SNR will
represent the ratio of the signal power to the variance of the
noise at the receiver of $D$.

\subsection{Existing cooperative diversity strategies}
We proceed to briefly introduce the above mentioned cooperative
diversity strategies.

\subsubsection{Non-dynamic selection-decode-and-forward}
In \cite{LanWorTSEIEEE}, Laneman, Tse and Wornell, described the
non-dynamic selection-decode-and-forward cooperation strategy for
the three-node network which asks for $R_2$ to cooperate, by first
decoding and then re-encoding, if and only if it is not in outage
with respect to the information source $S$.  The strategy requires
full channel knowledge at the receivers of relay $R_2$ and
destination $D$. Focusing on the case where each node has a single
transmit-receive antenna and where the fading scalars are from the
Rayleigh distribution, the scheme has $R_2$ decode if and only if
\begin{equation}\label{eq:decode iff g_2}
R_\text{source} < \log_2 (1 + \mbox{SNR} |g_2|^2)
\end{equation}
where $R_\text{source}$ is the information rate of the
transmission of $S$, measured in bits per channel use (bpcu).
 Again specific to the case of Rayleigh fading and of having a single transmit-receive
antenna per node operating in the half-duplex environment, it was
shown in \cite{LanWorTSEIEEE} that given a source-to-destination
information rate $R$, measured in bits per network channel use
(bpncu), and given a \emph{multiplexing gain} \cite{ZheTse}
$$r = \frac{R}{\log_2(\mbox{SNR})},$$ then the high-SNR
probability of network outage, and thus the optimal D-MG
performance, is given by:
$$ d_\text{ND-SDAFout}(r) := \frac{\log P_\text{outage}(r)}{\log\mbox{SNR}} = 2(1-2r), \ \ \ r_{\max}
= 1 / 2 $$ and that this optimal performance will be achieved by
some random Gaussian codes of infinite length. This result is
readily generalized to a network of $n+1$ users where the
cooperation strategy for intermediate relay $R_j$ is dictated by
\begin{equation}\label{eq:decode iff g_i}
R_j \in D(S) \ \Leftrightarrow \ R < \log_2 (1 + \mbox{SNR}
|g_j|^2), \ \ j=2,\cdots,n.
\end{equation}
Furthermore, given the implicit knowledge of the multiplexing gain
at the nodes\footnote{\label{foot:every node knows r }Knowledge of
the code/rate and of the power constraint, together with the
existing assumption of unit variance additive noise, jointly imply
knowledge of the multiplexing gain.}, and conditioned on infinite
time duration, the same approach translates to an optimal
performance of
\begin{equation} \label{eq:previous decode and forward performance}
d_\text{ND-SDAF-opt}(r) =
(n-1)\left(1-\frac{2n-1}{n-1}r\right)^{+}+(1-r)
\end{equation}
where $\alpha^{+}:= \max(0,\alpha)$.  The same performance was
predicted in \cite{LanWorTSEIEEE} for the non-dynamic amplify and
forward scheme.

Letting $r_{coop}$ describe the multiplexing gain corresponding to
the set of ($R,\text{SNR}$) pairs where the D-MG performance of
the cooperative scheme stops being equal to the D-MG performance
in the non-cooperative case, we note that given a rate $R$,
cooperation essentially applies only for SNR values greater than
$$\mbox{SNR} \ \dot \geq \ 2^{\frac{R}{r_{coop}}} =
2^{\frac{2n-1}{n-1}R}.$$
 As in \cite{ZheTse}, we have $\doteq, \ \dot\geq$ and $\dot\leq$
denoting asymptotic exponential equality and inequalities
respectively.
\begin{figure}[h]
\begin{center}
\begin{center}\includegraphics[angle = -90,width=0.7\columnwidth]{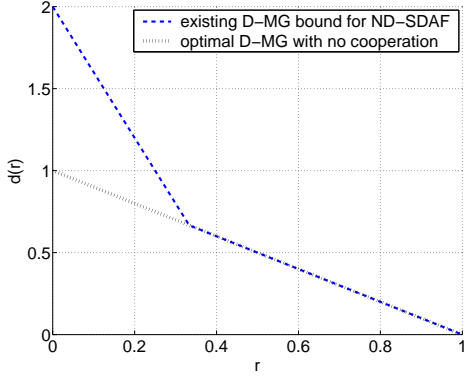}\end{center}
\caption{Existing D-MG bound for the single intermediate relay,
ND-SDAF strategy.  Given a network rate $R$ (bpncu), cooperation
applies for $\mbox{SNR} \ \dot \geq \ 2^{3R}$. \label{fig:previous
DMG bound for ND-SDAF}}
\end{center}
\end{figure}

\subsubsection{Non-dynamic receive-and-forward \label{sssec:RelayJournal introduction of ND-RAF}}
In a variant of the two-stage wireless relay network model
proposed by Jing and Hassibi in \cite{JingHass04b,JingHass04c}, we
will consider the case where the $n-1$ intermediate relays are
only allowed to perform linear-processing (time-averaging based on
space-time codes) on the received signal.  Knowledge of the
channel ($h_i,g_i$) is given only to the receiver of the final
destination $D$.

In \cite{JingHass04b,JingHass04c}, the authors also present a
bound on the network's pairwise error probability (PEP), which in
the high-SNR regime is a function only of SNR and of the minimum
eigenvalue of any difference of any two code matrices of the
corresponding distributed space-time code which performs the
linear-processing (see Appendix \ref{ssec:Appendix proof of
expanding old PEP}). This result, in conjunction with the finite
duration random coding proposed in \cite{JingHass04b}, guarantees
only for full-diversity but does not guarantee any bound on the
code eigenvalues.  As we discuss in Appendix \ref{ssec:Appendix
proof of expanding old PEP}, this implies that with finite delay
random coding, the maximum achieved multiplexing gain can get
arbitrarily small
\[d_\text{ND-RAF}(r) \geq (n-1)(1-kr)^{+}+(1-r), \ \ \ \ k>>1\] thus potentially requiring for asymptotically high SNR,
in order for cooperation to apply.

A better lower bound on the optimal D-MG performance of the
network is obtained when the same PEP result of
\cite{JingHass04b}, is applied on codes that are approximately
universal and thus D-MG optimal over any channel. In the same
Appendix \ref{ssec:Appendix proof of expanding old PEP}, we show
that given such codes, expanding the existing PEP bound towards
D-MG, guarantees a lower-bound on the network's D-MG performance
of:
\begin{equation}\label{eq:PEPbased_DMG}d_\text{ND-RAF}(r) \geq (n-1)\left(1-\frac{4n-1}{n-1}r\right)^{+}+(1-r).\end{equation} This implies a maximum diversity
${d(0) = n}$ and cooperation after
\begin{equation}\label{eq:SNRmin for ND-RAF from PEP}\mbox{SNR}_\text{min}
 \ \dot \geq \  2^{\frac{4n-1}{n-1}R}.\end{equation}
\begin{figure}[h]
\begin{center}
\begin{center}\includegraphics[angle = -90,width=0.7\columnwidth]{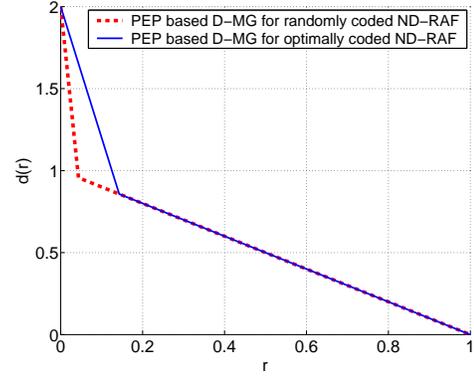}\end{center}
\caption{The PEP based D-MG bound for the single intermediate
relay ND-RAF strategy, predicts that given finite duration random
coding, cooperation applies for $\mbox{SNR} \ \dot \geq \ 2^{kR},
\ \ k>>1$ and that given optimal coding, cooperation applies for
$\mbox{SNR} \ \dot \geq \ 2^{7R}$. \label{fig:previously predicted
DMG performance for ND-RAF}}
\end{center}
\end{figure}

\subsubsection{Dynamic amplify-and-forward and dynamic receive-and-forward}

In this D-AAF strategy, first introduced in \cite{ElGamalDUplex},
the original source $S$ transmits at each time instance and each
intermediate relay takes turns in transmitting an amplified
version of a previously received signal, and do so only at even
time indexes. Again in \cite{ElGamalDUplex}, it is concluded that
for $n+1$ users, given infinite time duration, the optimal
tradeoff is given as
\begin{equation}\label{eq:optimal DMG performance D-AAF} d_\text{D-AAF}(r)
= (n-1)(1-2r)^{+}+(1-r).\end{equation} From the analysis in
\cite{ElGamalDUplex}, we can see that the same optimal D-MG result
holds when the intermediate relays do not use channel knowledge.

Having briefly introduced the cooperative strategies, we conclude
the introduction with a quick exposition of the basic tools to be
used in the network analysis and encoding.
\subsection{Preliminaries on the D-MG Tradeoff and approximate
universality} Let an $n\times T$ space-time code ${\cal X}$
operate at rate ${R_{\chi} = \frac{1}{T}\log_2(|\mathcal{X}|)}$
bpcu, and let $r_{\chi}$ be the multiplexing gain (normalized
rate) given by
$$r_{\chi} = \frac{R_{\chi}}{\log_2(\text{SNR})}$$ corresponding to
\begin{equation} \label{eq:card} | {\cal X} | \ = \
\text{SNR}^{r_{\chi}T}.
\end{equation} For large SNR, the capacity over an
$n\times n_r$ Rayleigh fading channel is given by $C_{\chi} \
\approx \ \min \{n,n_r \} \log_2(\mbox{SNR})$, implying a maximum
achievable multiplexing gain of ${r_{\chi,\max} = \min \{
n,n_r\}}$. The {\em diversity gain} corresponding to a given
$r_{\chi}$, is defined by
\[
d(r_{\chi}) \ = \ - \lim_{\text{SNR} \rightarrow \infty }
\frac{\log(P_{e})}{\log (\text{SNR})} ,
\]
where $P_{e}$ denotes the probability of codeword error. In a
recent landmark paper, Zheng and Tse \cite{ZheTse} showed that
there exists a fundamental tradeoff between diversity and
multiplexing gain, referred to as the diversity-multiplexing gain
(D-MG) tradeoff. For a fixed integer multiplexing gain $r_{\chi}$,
and $T\geq n+n_r-1$, the maximum achievable diversity gain
$d(r_{\chi})$ is shown to be
\begin{equation}
d(r_{\chi}) \ = \ (n-r_{\chi})(n_r-r_{\chi}) .
\label{eq:Zheng-Tse}
\end{equation}
The function for non-integral values is obtained through
straight-line interpolation. For $T < n+n_r-1$ only bounds on the
maximum possible $d(r_{\chi})$ were available \cite{ZheTse}.  It
was also shown in \cite{ZheTse} that there exist random Gaussian
codes with $T\geq n+n_r-1$ that achieve the above optimal
tradeoff. For such optimal codes, the probability of error
coincides with the probability of outage
\[P(\text{outage}) \ \dot = \ P(\text{error-optimal})\]
The Zheng-Tse result sparked considerable interest in meeting this
new D-MG frontier and explicitly providing D-MG optimal schemes.

The problem of constructing explicit D-MG optimal ST codes over
the Rayleigh-fading channel for any pair $(n,n_r)$ was settled in
\cite{EliRajPawKumLu} where it was shown that
cyclic-division-algebra-based space-time codes having a certain
non-vanishing determinant (NVD) property are optimal with respect
to the D-MG tradeoff of the Rayleigh-fading channel. The same
authors used this result to establish the D-MG optimality of
constructions of space-time codes found in Belfiore et. al.
\cite{BelRekVit} and Kiran and Rajan~\cite{KirRaj}. These prior
constructions were restrictive in terms of the values of $n$ that
can be accommodated.  In \cite{EliRajPawKumLu,isit05_explicit}, a
general construction of D-MG optimal codes was provided that was
valid for all $n,n_r$ and all $T\geq n$.

Furthermore, as is shown in \cite{TavVisUniversal_2005} (see also
\cite{EliaPerfect}), the above codes are approximately universal
\cite{TavVisUniversal_2005} and thus D-MG optimal over all
slow-fading channels, independent of the channels' statistics.  As
shown in the proof of Theorem 4.1 of \cite{TavVisUniversal_2005}
(Appendix A.2), due to the bound on their smallest eigenvalue,
such codes satisfy the extra property that
\begin{equation}\label{eq:approximately universal
optimality}P(\text{error} \ | \ \text{no outage}) \ \dot = \
e^{-\text{SNR}^{+}}  \ \dot = \ \text{SNR}^{-\infty} \ \dot = \
0.\end{equation} In a slight abuse of notation, we will use the
term `approximately universal given statistical symmetry' to
denote a D-MG optimal scheme that also satisfies
(\ref{eq:approximately universal optimality}), and does so for all
channel statistics with identically distributed path fading.

This concludes the introduction and we can now proceed with
analysis and encoding for the first cooperative strategy.

\section{Optimality in the Non-Dynamic Selection-Decode-and-Forward Strategy \label{sec:RelayJournal_ND-SDAF}}

In regards to the ND-SDAF strategy, we will here improve the
existing optimal performance bound in (\ref{eq:previous decode and
forward performance}), generalize to a new optimality expression
that holds for a larger family of channel statistics and
explicitly construct the first ever coding scheme that guarantees
for optimality in finite and minimum time delay, finite decoding
complexity and for any number of users.  We finally bound the
performance for a variety of network topologies and explicitly
provide encoding schemes that guarantee these bounds for any set
of statistics.
\paragraph{Cooperation strategy and general coding requirements}
We begin by noting that in this scheme, cooperation does not
provide gains unless the probability of erroneous decoding at the
intermediate relays is much smaller than the probability that some
relay does not select to cooperate. Necessary for D-MG optimality
is a decoding strategy that relates to outage as in
(\ref{eq:decode iff g_i}), in which case decoding at the node
(during the first stage) is required to provide for
\[P(\text{error at }R_i \ | \ \text{no outage}) << P(\text{error at $D$ with cooperation})\]
which in terms of approximate universality, translates to
\begin{equation}\label{eq:prob error given no outage is zero}P(\text{error at }R_i \ | \ \text{no outage}) \  \dot = \
\mbox{SNR}^{-\infty} \ \dot = \ 0.\end{equation} D-MG optimal
codes that guarantee that
\begin{equation}\label{eq:prob error is equal to the probability of outage}P(\text{error}) \  \dot = \
P(\text{outage})\end{equation} can be found in the family of
random Gaussian codes, which also maximize the mutual information
of both the first and second stages of the transmission, and hence
optimize the overall D-MG performance of an infinite duration
network. The lack of eigenvalue bounds in the finite length
version of these random codes does not allow for (\ref{eq:prob
error given no outage is zero}) to hold, and instead as shown in
\cite{ZheTse}, finite length random Gaussian codes satisfy,
\[P(\text{error} \ | \ \text{no outage}) \  \dot  =  \  P(\text{outage}) \ \dot > \
\mbox{SNR}^{-\infty}.\]  Consequently, the use of random Gaussian
codes in the first stage of the network transmission
(source-to-relay) immediately imposes a requirement for infinite
time duration \cite{LanemanII,LanWorTSEIEEE,ElGamalDUplex}.

We will proceed to explicitly provide a network encoding scheme
that meets the outage region of any statistically symmetric
ND-SDAF network and does so in finite time duration.
\subsection{Explicit D-MG optimal encoding for the statistically
symmetric ND-SDAF network} To achieve optimality, we jointly treat
the first and second stage encoding methods by first constructing
a $1\times n$ approximately universal code over SISO channels and
then an $n\times n$ approximately universal code over MISO
channels, which has the extra two properties that it maintains its
optimality even if it is truncated and that it has the optimal
number of common entries with the first stage SISO code.

\paragraph{First stage transmission and the horizontally-restricted,
approximately universal, perfect code} For a relay-network where
all the nodes have one transmit/receive antenna operating in
half-duplex over a statistically symmetric channel, the proposed
coding scheme asks for the source $S$ to sequentially transmit,
during time $t=1,2,\cdots,n$, the $n$-length vector $\underline{k}
= \theta \underline{z}$ where $\underline{z}$ comes from the
$1\times n$ \emph{horizontally-restricted perfect code}
$\mathcal{X}_h$,
\begin{eqnarray} \label{eq:horizontally restricted perfect code} \mathcal{X}_h & = & \{ \underline{z} = [ z^{(1)} \ z^{(2)} \
\cdots \ z^{(n)}] \}\\ & = & \{ [ f_1 \ f_2 \ \cdots \ f_{n}]M_n,
\ \ \forall [ f_1 \ f_2 \ \cdots \ f_{n}]\in
\mathcal{A}^n\}\nonumber\end{eqnarray} where the $f_i$ are from a
discrete information constellation $\mathcal{A}$ such as QAM or
HEX, that scales with SNR as $|\mathcal{A}| = \text{SNR}^r$, and
where $M_n$ is the unitary lattice generator matrix for perfect
codes \cite{PerfectCodes,EliaPerfect,EliaPerfectJournal}. Finally
$\theta$ is the normalization factor such that $\mathbb{E}[|\theta
z^{(i)}|^2] \dot = \text{SNR}$. The approximate universality of
$\mathcal{X}_h$ over the first stage SISO channel and the
subsequent satisfaction of (\ref{eq:prob error given no outage is
zero}), are established by observing that the code carries the
same information and has the same non-vanishing product distance
($\mathcal{X}_h$ consists entirely of one of the layers of the
CDA-perfect codes) as the simple QAM scheme, which was shown in
\cite[Section 3]{TavVisUniversal_2005} to be approximately
universal over all SISO channels.
\paragraph{Second stage transmission and the residual approximate
universality of CDA codes} By $t=n$, due to the approximate
universality of $\mathcal{X}_h$, each intermediate relay ${R_i \in
D(S)}$ has \emph{correctly} decoded $\underline{k}$ and will
participate in the second stage of cooperation which will take
place if and only if $r\leq \frac{1}{2}$ (this choice of $r$ will
become clearer later on), in which case the network encoding
scheme asks from each ${R_i\in D(S), \ i=2,\cdots,n}$ to transmit
$k^{(i)}$ (the $i^{th}$ element of $\underline{k}$) at time
$t=n+i-1$, thus allowing the decoder of $D$ to `see' a possibly
truncated version of the \emph{diagonal restricted perfect code},
$\mathcal{X}_d$, given as:
\begin{equation}\label{eq:diagonal_restricted_perfect_code dec&forw }\mathcal{X}_d = \{ diag(\underline{z})
= diag(\underline{f} \cdot M_n), \ \forall \underline{f}\in
\mathcal{A}^n\}.\end{equation} As with $\mathcal{X}_h$, the
corresponding information alphabet $\mathcal{A}$ is discrete and
$M_n$, as in (\ref{eq:horizontally restricted perfect code}),
represents the orthogonal matrix that generates a lattice in the
maximal field $\mathbb{L}$ of the division algebra of the perfect
codes (see equation (\ref{eq:G_Matrix})). The above choices make
$\mathcal{X}_d$ approximately universal over any i.i.d. MISO
channel, since the product-distance of any difference of diagonals
in the matrices of $\mathcal{X}_d$,
\begin{eqnarray*} \prod_{j=0}^{n-1}\delta z(j)  =
\prod_{j=0}^{n-1}\sigma^j\biggl(\sum_{k=0}^{n-1} f_k
\beta_k\biggr) =
N_{\mathbb{L}/\mathbb{Z}[\imath]}\biggl(\sum_{k=0}^{n-1} f_k
\beta_k\biggr) \in \mathbb{Z}[\imath] \end{eqnarray*} is an
algebraic norm and thus a Gaussian integer, meaning that
\[  \biggl| \ \prod_{j=0}^{n-1}\delta z(j) \ \biggr| \ \dot > \ 1, \] which together with the fact
that $|\mathcal{A}| = \mbox{SNR}^r$, satisfy all the related
conditions in \cite[Theorem 4.1]{TavVisUniversal_2005}.

The above approximate universality only relates to the event where
all intermediate relays $R_i$ are in $D(S)$.  As we have seen
though, it is the case that some relays might be unable to decode,
in which case the equivalent space-time code will not be the
complete $n\times n$ $\mathcal{X}_d$ but instead will be missing
some rows and will be of dimension $(|D(S)|+1)\times n$.  For
this, we now move to establish another property, necessary for the
network's approximate universality.  We will name this property as
\emph{`residual approximate universality'}.
\begin{defn}Residual approximate universality is the property of
an $n\times n$ approximately universal space-time code, which
guarantees that after removing an arbitrary number, say $k$, of
rows from each of the code matrices, it is then the case that the
resulting $(n-k)\times n$ truncated code is still approximately
universal over any $(n-k)\times n_r$ channel, for all
$n_r$.\end{defn} The proof that CDA-perfect codes are residually
D-MG optimal (and hence residually approximately universal) can be
found in the proof of \cite[Theorem 4]{isit05_explicit} and is
based on the fact that, given any truncated $(n-k)\times n$
codematrix $X$, the Hermitian nature of $XX^\dag$, guarantees that
the magnitude of each of its $n-k$ ordered eigenvalues, is each
lower bounded by the squared magnitude of the corresponding $n-k$
smallest eigenvalues of the original pre-truncated matrix.

A similar argument, gives that the above diagonal restricted
perfect code is also residually approximately universal, over any
MISO channel with i.i.d. fading.  Having constructed a proper pair
of residually approximately universal codes for the two stages, we
can state that:
\begin{thm}\label{thm:optimalDMG of decode and forward}
Given statistically symmetric Rayleigh fading, the optimal
half-duplex constrained D-MG performance of the non-dynamic
selection-decode-and-forward wireless network with $n+1$
single-antenna users, is given by:
\begin{eqnarray}d_\text{ND-SDAFopt}(r)  = (n-1)(1-2r)^{+}+(1-r).\end{eqnarray}
This performance can be achieved by utilizing a $1\times n$
horizontally-restricted perfect code during the first stage,
having the intermediate relays decode if and only if they are not
in outage with respect to the original transmitter, and finally
having them re-transmit utilizing, together with the source, a
distributed $n\times n$ diagonal-restricted perfect code.
\end{thm}
\begin{proof}
See Appendix \ref{ssec:proof of DMG optimal ND-SDAF}\end{proof}
\begin{figure}[h]
\begin{center}
\begin{center}\includegraphics[angle = -90,width=0.7\columnwidth]{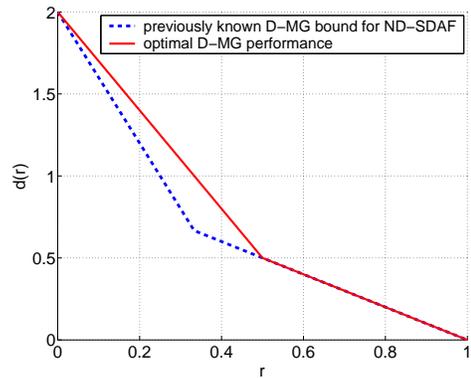}\end{center}
\caption{Optimal D-MG performance for the single-intermediate
relay, ND-SDAF scheme utilizing perfect codes. Given a rate $R$
(bpncu), cooperation applies for $\text{SNR} \ \dot \geq \
2^{2R}$. \label{fig:old and new DMG result in ND-SDAF}}
\end{center}
\end{figure}

For results relating to different channel statistics and different
channel topologies, we refer the reader to Appendix
\ref{ssec:Different channel statistics and topologies for
ND-SDAF}.

\section{Optimality in the Non-Dynamic Receive-and-Forward Strategy\label{sec:RelayJournal_ND-RAF}}
In this network, we initially consider the case where each node
has a single receive-transmit antenna operating in half-duplex.
The exact network introduced in Section \ref{sssec:RelayJournal
introduction of ND-RAF}, deviates from the relay setup in
\cite{JingHass04b,JingHass04c}, as it requires one less
intermediate relay since the destination $D$ does not discard the
direct transmission of $S$.  Unlike in the case of the ND-SDAF
strategy, the lack of information extraction at the relays,
introduces the need that the related space-time code be completely
distributable. Code distribution at the relays is discussed in
Appendix \ref{sec:Appendix_Perfect_Codes}.

Interest in this `linear-processing' (ND-RAF) relay network is
generated by the scheme's two main advantages. The first advantage
is that the intermediate relays do not require channel information
and the second advantage is that the intermediate relays do not
perform time and energy consuming decoding.  Potential power
violation issues arising from utilizing the above properties are
addressed in Sections \ref{sec:RelayJournal_D-RAF} and
\ref{sec:relay journal simulations}.

We proceed to further describe the relay network model and to
analyze the equivalent second stage `two-product' channel. The
main result will be presented in Theorem \ref{thm:optimalDMG of
ND-RAF}.
\subsection{Distributed space-time codes and the
equivalent channel}

\subsubsection{Relay scheme} During the first stage, the
source's single antenna sequentially transmits a vector
\begin{equation}\label{eq:Info_Vector_f}
{\underline{k} = \theta\underline{z} = \theta \cdot
\begin{array}{|cccc|} z^{(1)} &  z^{(2)} & \cdots & z^{(n)}
\end{array}}\end{equation}
of $n$ signals, where $\underline{z}$ is a codeword from a
$1\times n$ coding scheme. Each intermediate relay $R_i, \
i=2,3,\cdots,n$ then receives the $n$-length vector
\begin{equation}\label{received_vector_ri}\underline{r}_i = \theta g_i \underline{z}+\underline{v}_i,\end{equation}
independently performs linear-processing (time-averaging based on
space-time codes) on $\underline{r}_i$ and transmits
\begin{equation}\label{eq:relay_Tranmit_signal_xi}\underline{x}_i=
\underline{r}_i A_i\end{equation} where each $n\times n$ matrix
$A_i$ is unitary.

The signal at the receiver of the final destination $D$ is then of
the form
\begin{eqnarray} \label{eq:DistrST_model1}
\underline{y} & = & \sum_{i=1}^n h_i \underline{x}_i +
\underline{w}.
\end{eqnarray}
Clarifications regarding $h_1$ can be found in Appendix
\ref{ssec:jump to the two-product channel}.

\subsubsection{Equivalent channel model} The authors in
\cite{JingHass04b,JingHass04c} utilize a finite-length, random,
distributed space-time code by having the linear-processing at the
intermediate relays be performed by randomly chosen unitary
linear-dispersion matrices \cite{HassibiLDCs}, where each such
matrix uniquely defines a codematrix row. The rationale behind
this becomes clearer after rewriting (\ref{eq:DistrST_model1}) as:
\begin{eqnarray} \label{eq:DistrST_model2}
\underline{y} & = &  \sum_{i=1}^n h_i(\theta g_i
 \underline{z}+\underline{v}_i) A_i + \underline{w}\nonumber \\
& = & \theta \sum_{i=1}^n  h_ig_i \underline{z}A_i + \sum_{i=1}^n
h_i \underline{v}_i A_i + \underline{w}
\end{eqnarray}
which easily transforms to the familiar point-to-point channel
model
\begin{eqnarray} \label{eq:channelModel1}
\underline{y} & = & \theta HX+W
\end{eqnarray}
where
\begin{eqnarray} \label{eq:Channel_Matrix_Definition_1}
H &=& \begin{array}{|cccc|} g_1h_1 & g_2 h_2 & \cdots & g_n h_n
\end{array}\  \ ,  \ X = \begin{array}{|c|}
\underline{z}A_1\\
\underline{z}A_2\\
\vdots\\
\underline{z}A_n\\
\end{array}\\  W & = &  \sum_{i=1}^n h_i
\underline{v}_i A_i + \underline{w}. \nonumber
 \end{eqnarray}
Based on the results in Appendix \ref{ssec:jump to the two-product
channel} and for the sake of simplicity, we name the above channel
as the `two-product channel' by which we will exactly mean:
\begin{defn}\label{defin:definition of two product channel}The \emph{`two-product channel'} is the $n\times 1$
MISO channel, modelled as in (\ref{eq:channelModel1}), with the
channel matrix $H$ as in (\ref{eq:channelModel1}) representing
fading coefficients which are products of two i.i.d.
$\mathbb{C}\mathcal{N}(0,\mbox{SNR}^0)$ random variables.
Furthermore, the effective additive noise consists of spatially
and temporally white $\mathbb{C}\mathcal{N}(0,1)$ random
variables.
\end{defn}

\subsection{Outage probability of the two-product channel}
Due to the absence of decoding at the intermediate relays, the
rate-reliability limitations are mainly due to the
relays-to-destination stage of the network transmission. As a
result, the optimal D-MG tradeoff of the network is a function of
the half-duplex effect and of the optimal D-MG tradeoff of the
second stage two-product channel.

The general method for establishing the D-MG limits of the
two-product channel, relates to utilizing the existing eigenvalue
bounds of the approximately universal CDA codes, finding the
codes' `error contribution region' (channel region in which
perfect codes decode erroneously) and equating this region to the
outage region of the channel. Establishing the eigenvalue
statistics of the two-product channel then provides for the volume
of the outage region and for the optimality limits.  With respect
to these limits, we note that:
\begin{note}\label{note:two product no better than Rayleigh}
Since knowledge of $g_i$ at relay $R_i$ essentially reduces the
two-product channel to the Rayleigh fading channel, it is the case
that the two-product channel's optimal D-MG tradeoff cannot be any
better than that of the Rayleigh fading channel.
\end{note}
We proceed with the probability of outage for the two-product
channel.
\begin{prop} \label{prop:final_DMG_Of_Two_Product_Channel}
The optimal diversity-multiplexing gain tradeoff of the $n\times
1$ two-product channel is given by \[d_{\text{eq}}(r_\chi) =
n(1-r_\chi).\]
\end{prop}
\begin{proof}
From Lemma \ref{lem:Error_Contribution_Region} (Appendix
\ref{ssec:Appendix lemma on error contr region of perfect codes}),
we see that for $$ \lambda_n = HH^\dag=\sum_{i=1}^n
\|h_i\|^2\|g_i\|^2 := \mbox{SNR}^{-\mu} $$ being the only non-zero
eigenvalue of the two-product channel, then the corresponding
outage region, which at high SNR equals the error contribution
region of the perfect codes, is given by:
\begin{equation} \label{eq:region_Of_Error_Contribution}
\mathcal{B} = \{ \mu \geq 1-r_\chi\}.
\end{equation}

By definition of approximate universality, the probability of
error outside $ \mathcal{B} $ can be considered to be arbitrarily
small, allowing us to limit our attention only to the channels
with $ \lambda_n \, \dot \leq \, (\mbox{SNR}^0)$ \footnote{This is
immediate by first considering that for the Rayleigh fading case
$r_{\chi,\max} = 1$, and then by considering Remark \ref{note:two
product no better than Rayleigh}.}. As a result, knowledge of the
pdf of the channel outside this region is unnecessary, with the
only condition that $ {f_{\lambda_n}(\lambda_n)<\infty, \ \forall
\lambda_n \dot > \mbox{SNR}^0} $.  We will see later in Appendix
\ref{ssec:Appendix lemma on effective pdf of two-product channel}
that this condition is met. Immediately from Lemma
\ref{lem:Final_Channel_PDF} in Appendix \ref{ssec:Appendix lemma
on effective pdf of two-product channel}, we see that for
$h_i,g_i, \ i=1,2,\cdots,n$ being
$\mathbb{C}\mathcal{N}(0,\mbox{SNR}^0)$ random variables, then the
probability density function of $\lambda_n = \sum_{i=1}^n
\|h_i\|^2 \|g_i\|^2$ is upper bounded as
\begin{eqnarray}\label{eq:finalPDF}
f_\lambda(\lambda) \dot \leq \lambda^{n-1}
\end{eqnarray}
and for ${\lambda_n = \mbox{SNR}^{-\mu}}$
\begin{equation} \label{eq:finalSimplifiedPDF}f_{\mu}(\mu) \dot \leq \mbox{SNR}^{-\mu n}.\end{equation}

We here note that in \cite{ZheTse,EliRajPawKumLu}, the pdf of
$\mu$ for the $n\times 1$ Rayleigh fading channel, is given by
$${f_{\mu}(\mu) \dot = \mbox{SNR}^{-\mu n}e^{\mbox{SNR}^{-\mu}}}$$
which reduces to the two-product pdf expression of ${f_{\mu}(\mu)
\dot \leq \mbox{SNR}^{-\mu n}}$ for all ${\mu>1-r_\chi>0}$, that
is for all $\mu \in \mathcal{B}$. We then proceed as in
\cite{EliRajPawKumLu}, where $\mathcal{B}$ and $f_{\mu}(\mu)$
completely defined the probability of codeword error as
\[P_e\leq \int\limits_{\mu\in \mathcal{B}} f_{\mu}(\mu)
d\mu\] since the double exponential nature of the probability of
error, given a channel realization, acts as a binary indicator
function in and out of ${\mathcal{B}=\{ \mu
> 1-r_\chi \}}$.  As a result \[P_e\leq \int\limits_{\mu\in
\mathcal{B}} \mbox{SNR}^{-\mu n} d\mu\] and using Varadhan's Lemma
\cite{DemZei} or the dominant term approach of Appendix II of
\cite{EliRajPawKumLu}, we get that
\begin{equation}\label{eq:Varadhans_Resulting_Exponent}P_e  \ \dot \leq \  \max\limits_{\mu \in \mathcal{B}} \{
\mbox{SNR}^{-\mu n} \}.\end{equation} For a given multiplexing
gain, the maximizing eigenvalue is then ${\mu = 1-r_\chi}$, which
implies that
\[P_e  \ \dot \leq  \ \mbox{SNR}^{-n(1-r_\chi)}\]

This lower bounds the optimal D-MG tradeoff in the $n\times 1$,
equivalent two-product channel as
\[ d_{\text{eq}}(r_\chi)\geq n(1-r_\chi).\]
Remark \ref{note:two product no better than Rayleigh} completes
the proof. \end{proof}

We are now in position to give an explicit description of a D-MG
optimal coding method for the ND-RAF network where the
statistically symmetric Rayleigh fading channel is known only at
the receiver of the final destination $D$, and where each node has
a single transmit-receive antenna operating under the half-duplex
constraint.  The rate and the power constraint are known at all
nodes.

\begin{thm}\label{thm:optimalDMG of ND-RAF} (D-MG optimality in
the ND-AAF and ND-RAF):  The optimal half-duplex constrained D-MG
performance of the ND-RAF network with $n+1$ single-antenna users,
is given by:
\begin{equation} \label{eq:optimal DMG for ND-RAF}
d_\text{ND-RAFopt}(r) = (n-1)(1-2r)^{+}+(1-r).
\end{equation}
The performance is achieved by utilizing an approximately
universal $1\times n$ horizontally-restricted perfect code during
the first stage $t=1,2,\cdots,n$.  If $r\geq \frac{1}{2}$ then the
relays do not forward the message and the source begins with the
new message at $t = n+1$.  If $r<\frac{1}{2}$ then each
intermediate relay $R_i, \ i=2,\cdots,n$ forwards at time
$t=i+n-1$ what it received at time $t = i$. Decoding at the final
destination uses the received signals at either time slots $t =
1,2,\cdots,n$ ($r\geq \frac{1}{2}$), or at time slots $t =
1,n,n+1,\cdots,2n-1$ ($r< \frac{1}{2}$).
\end{thm}
\begin{proof}
For $r>\frac{1}{2}$, the result is immediate by observing that the
equivalent channel is a SISO channel and the equivalent code is
the approximately universal horizontally-restricted perfect code
(\ref{eq:horizontally restricted perfect code}). For $r\leq
\frac{1}{2}$, the equivalent channel is the two-product channel,
and the equivalent space-time code is the $n\times n$
diagonal-restricted perfect code
(\ref{eq:diagonal_restricted_perfect_code dec&forw }) that is D-MG
optimal over all statistically symmetric channels. Consequently
the proof is immediate through Proposition
\ref{prop:final_DMG_Of_Two_Product_Channel} and from the fact that
transmission takes place during $2n-1$ time slots, shown in the
proof of Theorem \ref{thm:optimalDMG of decode and forward} to be
the minimum allowed.
\end{proof}
\begin{figure}[h]
\begin{center}
\begin{center}\includegraphics[angle = -90,width=0.7\columnwidth]{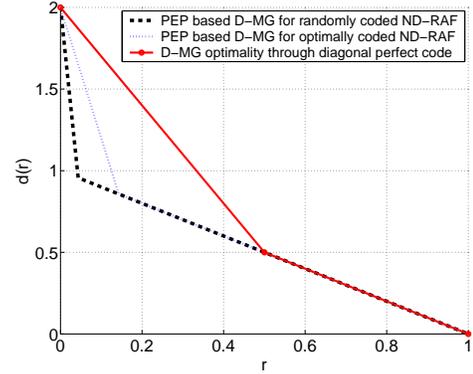}\end{center}
\caption{D-MG optimal performance for the single intermediate
relay, ND-RAF scheme vs the PEP based D-MG bounds for random and
optimal coding. \label{fig:DMG optimal ND-RAF vs
previouslypredicted DMGs}}
\end{center}
\end{figure}
We now proceed to provide the general coding requirements for
optimality in the ND-AAF and ND-RAF schemes, and to give bounds on
the performance of networks that utilize some other existing
coding methods.

\subsection{General coding optimality conditions in the ND-RAF relay network}
A closer look at the structure of standard perfect codes carrying
$n^2$ information elements, reveals that approximate universality
is a necessary but not a sufficient condition for network
optimality, and that these standard-perfect codes do not provide
for ND-RAF optimality as they fail to address the extra
restrictions introduced by the half-duplex constraint. These extra
constraints, exposed here in order to provide guidelines for
future encoding methods, are a direct consequence of the fact that
one cannot extract information at the intermediate relays, and can
manipulate the received signals only with linear transformations.
\paragraph{Condition 1:  The
distributed code must have at least one row that is the ordered
subset of the vector transmitted by the source} This will allow
for the destination's decoder to utilize the signal of $S$.
\paragraph{Condition 2:  The $n\times T$ equivalent code should map exactly $T$ information symbols from a discrete alphabet}
Specific to the minimum delay case, we see in Lemma
\ref{lem:Rate_Drop_Due_To_Duplex} of Appendix
\ref{ssec:Rate_Reduction_Due_To_Duplex}, that an $n\times T$
space-time code, carrying $mT$ information elements from a
discrete constellation and operating (in the point-to-point sense)
at multiplexing gain $r_\chi$, will allow for the network to
operate at multiplexing gain $\frac{r_\chi}{m+1}$. On the other
end of the spectrum, Proposition
\ref{prop:mn_discrete_symbols_rmax_is m} tells us that the same
code operating in the two-product channel with $n$-transmit
antennas ($m \leq 1$), will achieve maximum multiplexing gain of
$r_{\chi,\max} \leq m$.   Consequently, optimal performance
requires $m=1$.

Standard perfect codes, with $m=n$, fail this condition and
provide for
$$d_{\text{network-full}}(r) = (n-1)\left(1-\frac{n^2+n-1}{n-1}r\right)^{+} + (1-r)$$
guaranteeing for commencement of cooperation, given some rate $R$,
only after
$$\mbox{SNR}_\text{coop-full} \ \dot \geq \  2^{\frac{n^2+n-1}{n-1}R}$$
making it so that each new user entry renders the network less
cooperative.

\paragraph{Condition 3:  Code does not require complex-conjucacy}
The fact that complex-conjucacy is not a matrix operation, forces
the source to send twice as many real information elements, thus
doubling the duration of the first stage.

For the single intermediate relay case, the Alamouti code, which
can be shown to be optimal over the two-product channel, only
provides for $$d_{\text{network-alam}}(r) = (1-5r)^{+} + (1-r)$$
guaranteeing for commencement of cooperation only after
$$\mbox{SNR}_\text{coop-alam} \ \dot \geq \  2^{5R},$$ in contrast
to the optimal case, given by the diagonal-restriction perfect
code of $\mbox{SNR}_\text{coop-opt} \ \dot \geq \  2^{2R}$.

For high values of $n$, orthogonal designs are expected to carry
on the average $m \approx \frac{1}{2}$ information symbols per
channel use and be of dimension $n\times n^2$. Analysis similar as
that presented in the proof of Lemma
\ref{lem:Error_Contribution_Region} in Appendix \ref{ssec:Appendix
lemma on error contr region of perfect codes} (also see
\cite{EliKumPawRajkRajLu}), gives that the D-MG performance in the
second stage is given by $d(r) = n(1-\frac{r}{m})\approx n(1-2r).$
The first stage duration needs to be of length $2n^2m \approx n^2$
since conjugation cannot be described as a matrix transformation,
providing for $$d_{\text{network-ortho}}(r) \approx
(n-1)(1-7r)^{+}+(1-r),$$
 guaranteeing for
commencement of cooperation only after
$$\mbox{SNR}_\text{coop-ortho} \ \dot \gtrsim \  2^{7R}.$$

\subsection{Coding method for a network with a large number of users - the integral restriction perfect code}
We here present the `\textit{integral-restriction perfect code}'
which manages to exhibit excellent performance especially as the
number of users increases. Given a discrete information set
$\mathcal{A}$, the code is given by:
\begin{center}$\mathcal{X}_{ir} = \{ X =
\sum_{k=0}^{n-1} f_k \Gamma^k, \ \forall f_k\in \mathcal{A} \}
$\end{center} where $\Gamma$ from (\ref{eq:GammaMatrix}) provides
the linear-dispersion matrices $A_i = \Gamma^{n-i}, \ \
i=1,\cdots,n$. In addition to satisfying the necessary conditions
of:
\begin{enumerate} \item \emph{raw data in one row} \item \emph{one discrete
information symbol per channel use} \item \emph{no complex
conjucacy}\\

\hspace{-21pt}it also manages to have:\\

\item \emph{Constellation complexity remains constant as the
number of users increases:} Due to the special nature of $\Gamma$
(see \cite{EliaPerfect,EliaPerfectJournal}) and after a small
modification, the transmitted signals essentially belong in a
normalized QAM-HEX signalling set, independent of the number of
users. \item \emph{Fast encoding at the intermediate relays:} Due
to the sparse and discrete nature of the $A_i$'s, the scheme only
requires one multiplication with a small Gaussian integer, per
channel use. \item \emph{Network topology translates to reduction
of the sphere-decoding complexity at the destination:} The
existence of only one receive antenna at the destination
translates into a sphere-decoding complexity reduction at the
final destination receiver, from $O(n^2)$ to $O(n)$. \item
\emph{Allows for optimal rate, ease of construction and minimum
delay, for any $n$}.
\end{enumerate}

\subsection{Communication with a base-station: reducing the half-duplex effect} We will see that
an increase in the number of antennas at the destination
(base-station) does not only increase the diversity gain, given
cooperation, but it also reduces the SNR required for cooperation
to commence.  This is done at the cost of extra decoding
complexity at the final destination but bares no cost for the
relays.

In more detail, we recall a practical relay-network scenario given
in \cite{LanWorTSEIEEE} that talks of several relays with one
transmit/receive antenna, cooperating in their task to communicate
with a single base-station.  It is logical to assume that the
centrality of such a base-station will allow it to utilize
multiple receive antennas. This can correspond to a wireless
telephony setup where each mobile user utilizes the surrounding
users to increase the reliability of the transmission to the
base-station.

It should be noted that the fact that the relays can only perform
linear processing, prohibits having multiple antennas at the
source since one cannot linearly process matrices in a meaningful
way due to the additive nature of the received signal at the
intermediate relays. Furthermore the intermediate relays can only
have one receive-transmit antenna due to the lack of
source-to-relay channel information.

The following theorem explains how, given some network rate $R$,
having multiple receive antennas at the base station can allow for
a relay network with single-antenna intermediate relays to
potentially reduce the SNR required for cooperation to apply.
\begin{prop} \label{prop:rate reduction in base station}
Consider a base-station centered, ND-RAF relay network of $n$
users cooperating through an $n\times n$ space-time code whose
D-MG performance over the equivalent (second-stage) channel is
$d_{\text{eq}}(r)$. Let the code map on the average $mn$
information elements.  If the base station utilizes $m$ receive
antennas, $1 < m \leq n$, then cooperation is beneficial for
multiplexing gains that are smaller than the multiplexing gain at
the intersection of curves $d_{\text{eq}}(r(m+1))$ and  $d(r) =
m(1-r)$.
\end{prop}
\begin{proof} Since the information constellation satisfies $|\mathcal{A}|
\dot \leq \mbox{SNR}^\frac{r}{m}$ and since the entire network
transmission has duration $mn+n$ time slots, then under forced
cooperation we have that
\[r  =\frac{ R_{\text{network}}}{\log_2(\mbox{SNR})} = \frac{\frac{mn}{mn+n}\log_2(\mbox{SNR}^\frac{r_{\chi}}{m})}{\log_2(\mbox{SNR})} =
\frac{r_{\chi}}{m+1}.\]
\end{proof}
\begin{ex}For $n=4$ users cooperating to communicate with a
base-station, doubling the number of antennas at the base station
from $m=1$ to $m=2$ will allow for the necessary SNR, for
cooperation to apply, to be reduced from $\mbox{SNR}\geq 2^{2R}$
to $\mbox{SNR}\geq 2^{\frac{5}{3}R}$.\end{ex} The above bound
hints towards utilizing $n\times n$ codes that map $mn$
information elements and maintain sufficiently good eigenvalue
bounds for increasing spectral efficiency.  For this we turn again
to the general family of CDA/perfect codes and consider the
`$m$-layered $n\times n$ perfect code' $\mathcal{X}_m$, given by:
\begin{center} $\mathcal{X}_m = \biggl\{X = \sum_{j=0}^{m-1} \Gamma^j \biggl( diag \bigl(
\underline{f}_{j}\cdot G \bigr) \biggr)\biggr\}$
\end{center}
which maps the $mn$ information symbols from
$\{\underline{f}_0,\underline{f}_1,\cdots,\underline{f}_{m-1} \}$.
The codes have not been proven to be approximately universal.

We conclude that one could accept an increase in decoding
complexity and equipment, both only at the base station, in order
to save power at the intermediate relays and to increase the SNR
range in which cooperation is meaningful.

We now move to the dynamic receive-and-forward cooperative
diversity strategy.

\section{Optimality in the Dynamic Receive-and-Forward
Strategy\label{sec:RelayJournal_D-RAF}}
\subsection{Describing the scheme's model} For completeness, we will here reproduce the
description and outage analysis of the dynamic amplify-and-forward
network, first presented in \cite{ElGamalDUplex} for infinite time
duration, and we will then proceed to explicitly achieve this
optimality in finite time duration.

According to this strategy, the original source $S$ transmits at
each time instance, and each intermediate relay takes turns in
transmitting an amplified version of a previously received signal.
During a $2(n-1)$-length frame, all intermediate relays have
contributed, and the frame is repeated infinite times. The set of
equations that describe each frame, as given in
\cite{ElGamalDUplex}, is
\begin{eqnarray*} \label{eq:amp and for set of equations1b}
y_t & = & h_1 x_t + w_t , \ \ t-\text{odd}\\
y_t & = & b_ih_i(g_ix_1+v_{i,1})+h_1x_t+w_t , \ \
\begin{array}{c}i=\frac{t}{2}+1\\t-\text{even}\end{array}\nonumber
\end{eqnarray*}
or equivalently
\begin{eqnarray} \label{eq:amp and for set of
equations1}
y_t & = & h_1 x_t + w_t , \ \ t-\text{odd}\\
y_t & = & b_ih_ig_ix_1+h_1x_t+b_ih_iv_{i,1}+w_t , \ \
\begin{array}{c}i=\frac{t}{2}+1\\t-\text{even}\end{array}\nonumber
\end{eqnarray}
where $b_i$ is the amplification factor at relay $R_i$.  For
D-RAF, we set $b_i = 1$, which, given a minimum and usually very
small SNR, will always result in reduced average power consumption
and thus no power violation occurs at the cooperating relays.

\begin{prop}\cite{ElGamalDUplex}
\label{prop:multiple inter relay main result of Azarian D-RAF}
Given infinite time duration, the optimal D-MG performance of the
D-AAF and D-RAF schemes with $n-1$ intermediate relays, is given
by:
$$d_{\text{D-RAF}} = (n-1)(1-2r)^+ + (1-r).$$
\end{prop}
\begin{proof}For completeness, the proof in \cite{ElGamalDUplex} is briefly reproduced in Appendix \ref{ssec:appendix proof of
n=2 D-AAF Azarian}
\end{proof}
\subsection{Optimal explicit construction for the dynamic receive-and-forward
network} We will here present an explicit construction, based on
vectorized perfect codes, which achieves the optimal D-MG
performance of the dynamic-receive-and-forward scheme and does so
by using $2(n-1)$ frames, each of duration $2(n-1)$.
\begin{thm}\label{thm:D AAF all relays}
We consider the scheme where the source transmits continuously
from a set $x_1,x_2,\cdots,x_{4(n-1)^2}$ coming from a
column-by-column vectorization of a $2(n-1)\times 2(n-1)$ D-MG
optimal CDA space-time code, and where each intermediate relay
$R_i$, $i=2,\cdots,n$ forwards at time ${t = 2(i-1)+2(n-1)k, \
k=0,1,\cdots,2(n-1)-1}$ what it received at time ${t =
1+2(n-1)k}$.  This $n-1$ intermediate relay D-RAF scheme achieves
the optimal D-MG performance, as described in
\cite{ElGamalDUplex}, of
$$d_\text{D-RAF}(r) = (n-1)(1-2r)^+ + (1-r).$$
\end{thm}$ \\[-4pt] $
\begin{proof} See Appendix \ref{ssec:appendix proof n=2 vectorized perfect
D-RAF}.\end{proof}
$ \\[-10pt] $
The above construction is optimal for any number $n$ and for any
statistical asymmetry. We here note that in
\cite{Belfiore_Dynamic_AAF_allerton05}, a CDA-based D-RAF scheme
is proposed that, given statistical symmetry, achieves the optimal
tradeoff with delay $T = 4(n-1)$.
\subsubsection{The emerging need for high-dimensional perfect
codes} We have seen that in non-dynamic cooperative schemes,
operating below some SNR threshold $\mbox{SNR}_\text{coop}  \ \dot
= \ 2^{ \frac{R}{r_{\text{coop}   }   }}$ will prompt the network
to instruct the relays to not cooperate.  If one insists on
relaying even if $\mbox{SNR}<\mbox{SNR}_\text{coop}$, then through
CDA-based D-RAF, the same non-cooperative D-MG performance is
expected (at higher decoding and signalling complexity).  At low
rates and high multiplexing gains, the corresponding SNR is bound
to be small, rendering D-MG inaccurate and bringing us closer to
outage capacity which relates to low SNR and which accentuates the
role of maximization of mutual information. For a fixed rate, as
the SNR increases, approximate universality becomes more accurate
and will guarantee for the fidelity of the results in Theorems
\ref{thm:optimalDMG of decode and forward} to \ref{thm:D AAF all
relays}. This high-low SNR duality emphasizes the need for
approximately universal codes that are also information lossless,
offer Gaussian-like signalling and tend to maximize mutual
information, i.e. perfect space-time codes. This dual requirement,
together with the fact that in D-RAF the dimensionality of the
corresponding codes grows fast with the number of users, jointly
offer a substantial reason of existence for high-dimensional
perfect codes \cite{EliaPerfect,EliaPerfectJournal}.

\section{Comparing the presented cooperative diversity strategies\label{sec:RelayJournal_Comparison of schemes}}
A direct consequence of Theorems \ref{thm:optimalDMG of decode and
forward}, \ref{thm:optimalDMG of ND-RAF} and \ref{thm:D AAF all
relays}, is that:
\begin{thm}\label{thm:all schemes are the same}
The non-dynamic selection-decode-and-forward, the non-dynamic
receive-and-forward, the non-dynamic amplify-and-forward, the
dynamic amplify-and-forward and the dynamic receive-and-forward
cooperative diversity strategies, provide for the same D-MG
optimal performance.  For all the above cooperative strategies, a
network of $n+1$ users, each having knowledge of the rate and
power constraint, and each having a single transmit-receive
antenna operating in half-duplex over Rayleigh fading, has a
high-SNR probability of outage (optimal D-MG performance)
$P_\text{out} := \mbox{SNR}^{-d_\text{opt}(r)}$, where:
\begin{eqnarray}d_\text{opt}(r)  = (n-1)(1-2r)^{+}+(1-r).\end{eqnarray}
Optimality is achieved in minimum delay.
\end{thm}
\begin{figure}[h]
\begin{center}
\begin{center}\includegraphics[angle = -90,width=0.7\columnwidth]{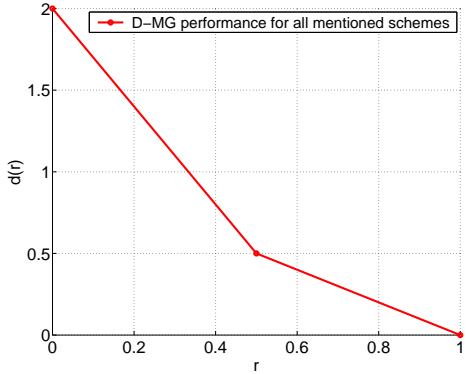}\end{center}
\caption{D-MG performance of the 3-node network for the above
cooperative diversity strategies (ND-SDAF, ND-AAF, ND-RAF, D-AAF,
D-RAF). \label{fig:D-MG performance for all mentioned schemes}}
\end{center}
\end{figure}

\section{Simulations \label{sec:relay journal simulations}}
In Figure \ref{fig:performance comparison of all coop-diversity
schemes} we present a comparison of the performance of all
mentioned cooperative diversity schemes.
\begin{figure}[h]
\begin{center}
\includegraphics[angle=-90,width=0.7\columnwidth]{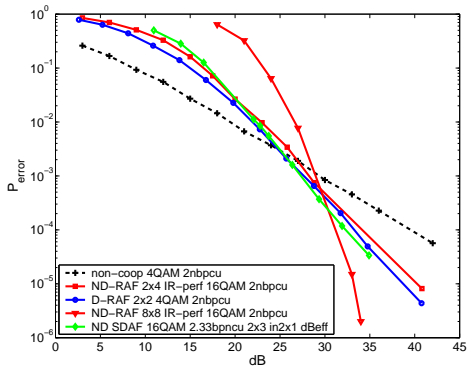}
\caption{Performance comparison for the 3-node ($n=2$) D-MG
optimal ND-SDAF, ND-RAF, D-RAF cooperative diversity schemes. The
schemes are also compared with the 9-node ND-SDAF and the
non-cooperative case.  Network operates at $2$ and $2.33$ nbpcu.
 \label{fig:performance
comparison of all coop-diversity schemes}}
\end{center}
\end{figure}
The ND-AAF scheme, in which the amplification factor is based on
the channel and set to equate the power consumption over the
different frequencies, constantly performs worse than all other
mentioned cooperation schemes of equal dimension.

The plots assume that the users are asked to cooperate at any SNR.
For the D-RAF, the amplification factor $b_2$ was set to $b_2 =
1$, with the condition that the transmission over the second
frequency (second row) did not violate the power constraint.  This
is always the case for the $\text{SNR}$ range of comparison.  The
SNR recorded is the SNR representing the total power used, given
that $b_2 = 1$. The same is valid for the ND-RAF case.  The code
used in the D-RAF case was the vectorized $2\times 2$ perfect code
operating over $4$-QAM ($M^2 = 4$). In the $n=2$ ND-SDAF, the code
used was the $2\times 3$ version of the diagonal restricted code,
operating over $16$-QAM, and in the $n=2$ ND-RAF, the code used
was the $2\times 4$ version of the diagonal restricted code, again
operating over $16$-QAM. We note that this choice is not optimal
but was used for reasons of network-rate uniformity.  No
information was allowed to be wasted at the receiver. In the $n=8$
ND-RAF, the code used was the $8\times 8$ integral-restriction
perfect code. The decoding strategy of the first stage of the
ND-SDAF network was such that the first stage
horizontally-restricted code did not limit the error performance
of the scheme. This was ensured by allowing the relay to decode
only if
 $$ |g_2|^2 >\delta\frac{M^2-1}{\text{SNR}}$$ where $\delta$ is chosen so
 that the effective rate is not the maximum allowed $8/3$ bpncu, but instead $7/3 =
 2.33$ bpncu. This small drop in rate allowed for decoding to take place well
 out of outage, thus guaranteeing that the probability of error at
 the relay (given a horizontally-restricted perfect code and a given
$\delta$) is less than the probability of error at the
 receiver of the destination, given that cooperation took place.
The choice of $\delta$ and of the corresponding effective rates
were arbitrary and somewhat heuristic.  The following figure gives
us an idea of the different choices and their utility.
\begin{figure}[h]
\begin{center}
\includegraphics[angle=-90,width=0.7\columnwidth]{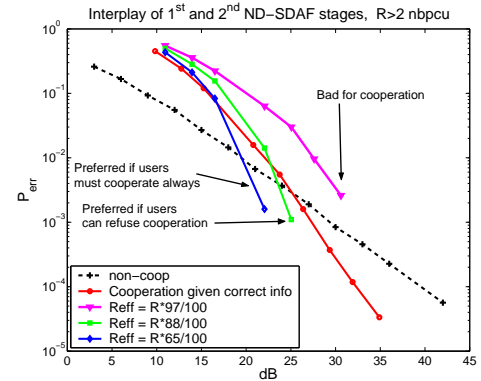}
\caption{Different decoding strategies at the intermediate relay,
and their effect as to how fast cooperation is allowed to be
fruitful. \label{fig:different deltas and Reff}}
\end{center}
\end{figure}
What we can conclude is that if the users have the choice to
accept or reject cooperation then the preferred decoding strategy
should be the one that allows for the maximum effective rate and
intersects the performance of the cooperation scheme before the
performance of the cooperation scheme intersects the performance
of the non-cooperative case.  On the other hand, if the users are
forced to cooperate independent of SNR and rate, then the
preferred decoding strategy presents a tradeoff between the
effective rate and the SNR required for the probability of error
at the intermediate relay to become smaller than the probability
of error given cooperation.

\section{Conclusion\label{sec:RelayJournal_Conclusion}}
 In this work we explicitly provided the first ever D-MG optimal
 encoding schemes for several cooperative diversity strategies for wireless relay
networks.  These practical perfect-code based schemes are optimal
over a broad range of channel statistics and topologies.  Bringing
the results together we were able to conclude that the ND-SDAF,
the ND-RAF (ND-AAF) and the D-RAF (D-AAF) have exactly the same
D-MG optimal tradeoff.

We proceed to recap the findings that relate to the different
schemes.

\bit
    \item
    For statistically symmetric ND-SDAF, ND-AAF and ND-RAF, optimality is provided by
    combining the horizontally-restricted perfect code over the first network stage, and the diagonally
    restricted perfect code over the second stage.  This
    optimality holds for any i.i.d. fading, any number of users and with the least possible delay
    of $T=1, T=n$ or $T = 2n-1$.

    \item
    Optimality in D-AAF and D-RAF is achieved by letting the source transmit continuously
    from a set $x_1,x_2,\cdots,x_{4(n-1)^2}$ coming from a
    column-by-column vectorization of a $2(n-1)\times 2(n-1)$ D-MG
    optimal CDA space-time code, and where each intermediate relay
    $R_i$, $i=2,\cdots,n$ forwards at time ${t = 2(i-1)+2(n-1)k, \
    k=0,1,\cdots,2(n-1)-1}$ what it received at time ${t =
    1+2(n-1)k}$.  Optimality holds for any number of users and any
    statistical distribution of the fading.
    \item
    For D-MG optimality, in a finite time duration ND-SDAF network, the
    first stage code needs to satisfy $P(\text{err} \ | \ \text{ no
    outage})<< P(\text{outage})$, and the second stage code needs to
    maintain (the corresponding) D-MG optimality even when any number
    of its rows are deleted.
    \bit
        \item It was noted that D-MG optimal random Gaussian codes do not
        satisfy the first stage condition unless they are of infinite
        length.
    \eit
\eit

In comparing the ND-RAF, in a statistically symmetric network, with the other schemes we see that it :\\
\begin{itemize} \item requires no channel knowledge at the intermediate relays
(compared to full knowledge for the ND-SDAF) \item has minimum
delay of $1$ or $2n-1$ (compared to $T = 4(n-1)$ of the
least-delay D-AAF scheme \cite{Belfiore_Dynamic_AAF_allerton05})
\item has signalling complexity of $n$ (compared to complexity of
$2(n-1)$ for the D-AAF scheme) \item has sphere decoding
complexity of $n$ (compared to ${4(n-1)}$ of the best proposed
D-AAF scheme).
\end{itemize}
As a final thought, it is pointed out that in terms of D-MG, what
was believed to require complete channel knowledge at the
receivers of the intermediate relays, require encoding over
infinite time duration and require decoding of infinite
complexity, was exactly achieved by newly constructed perfect code
variants, given absolutely no channel knowledge, given minimum
time duration and small decoding complexity.

\appendices
\section{Results and proofs relating to the ND-SDAF scheme
\label{sec:Appendix_proofs for ND-SDAF}}

\subsection{Proof of Theorem \ref{thm:optimalDMG of decode and forward} \label{ssec:proof of DMG optimal ND-SDAF}}
Given the approximately universal $\mathcal{X}_h$ and the
residually approximately universal $\mathcal{X}_d$, it is the case
that the network's full-duplex D-MG performance will be optimal
over any channel with i.i.d fading, and will be given as,
\begin{eqnarray*}P_\text{err} & = & P_{\mathcal{X}_h}(1,1)\biggl(
\sum_{i=0}^{n-1} P(|D(r_0)| = i)P_{\mathcal{X}_d}(i,1) \biggr)\\ &
\dot = & P_{\text{out}}(1,1)\biggl( \sum_{i=0}^{n-1} P(|D(r_0)| =
i)P_{\text{out}}(i,1) \biggr)\end{eqnarray*} even for the minimum
delay case, where $P_{\text{out}}(k,l)$ represents the probability
of outage in the $k\times l$ channel. For the Rayleigh fading MISO
channel, we have that
\[P_{\text{out}}(i,1) = (P_{\text{out}}(1,1))^i \ \dot = \  \mbox{SNR}^{-i(1-r)}\] resulting in
the full-duplex probability of network transmission error, for the
statistically symmetric Rayleigh fading case, to be given by
\begin{eqnarray*}
P_{\text{err}} &  = & P_{\text{out}}(1,1)\biggl( \sum_{i=0}^{n-1}
(P_{\text{out}}(1,1))^{n-1-i} (P_{\text{out}}(1,1))^i   \biggr) \\
& \dot = &  (P_{\text{out}}(1,1))^n \ \dot = \
\mbox{SNR}^{-n(1-r)}.
\end{eqnarray*}
In regards to the half-duplex constrained D-MG performance in the
statistically symmetric case, we observe that the overall network
transmission duration was $2n-1$ time slots.  The task is now to
prove that the duration cannot be any less.  For this we first
turn to Proposition \ref{prop:mn_discrete_symbols_rmax_is m} which
tells us that to achieve a maximum (second stage - MISO)
multiplexing gain of $r_{\max} = 1$, we need to map $n$ discrete
symbols in $\mathcal{X}_d$.  For this we need at least $n$ time
slots in the first stage (SISO) since any smaller duration would
violate the $r_{\max} = 1$ limit placed by the first stage SISO
channel.  The proof is then complete by observing that we need at
least one time slot for every row of $\mathcal{X}_d$ that does not
correspond to direct transmission.

\subsection{D-MG tradeoff for different fading
distributions and different channel topologies
\label{ssec:Different channel statistics and topologies for
ND-SDAF}}
\paragraph{Optimal D-MG performance for some other fading
distributions} Due to statistical symmetry, it will be the case
that, for any channel probability density function (pdf), the
network's D-MG optimality requires the same outage-based decoding
strategy (\ref{eq:decode iff g_i}), as well as the joint element
property and the residual approximate universality of the utilized
space-time codes. For an exact network performance expression, we
also need to know the codes' D-MG performance over the
corresponding channels. Deviating from Rayleigh fading, analysis
in \cite[Prop 5.2]{TavVisUniversal_2005} provides the optimal
tradeoff curve for the class of $n\times 1$ MISO channels with
i.i.d fading coefficients $\{c_i\}_{i=1}^n$, to be
\[d^{*}(r_\chi) = \alpha n(1-r_\chi), \ \ \ \ \ \ 0\leq r_\chi \leq 1\]
where $ \alpha := \lim\limits_{t\rightarrow 0} \frac{\log
\bigl(P(\|c_i\|^2 \leq t)\bigr)}{\log(t)} $.  We conclude that the
overall network half-duplex D-MG optimal performance in such a
channel is
\begin{eqnarray*}
d_\text{ND-SDAFopt}(r)  = \alpha(n-1)(1-2r)^{+}+\alpha(1-r).
\end{eqnarray*}

\subsubsection{Encoding schemes for networks with nodes having
multiple receive and transmit antennas} We consider the special
case where each terminal has $m$ receive-transmit antennas, and
where the fading across the different antennas has an
\emph{arbitrary statistical distribution}, with this distribution
being the same over any path.  This topology brings to the fore
the \emph{'horizontally-stacked perfect code'} and the following
bound.
\begin{prop}\label{prop:horizontally stacked optimality}
The full-duplex D-MG performance of the ND-SDAF with $n+1$ nodes,
each having $m$ transmit-receive antennas, is given by the
integral-point-wise linear plot governed by:\[ d^{*}(r_\chi) =
n(m-r_\chi)^2 \ , \ \ \ \ \ r_{\chi,\max} = m\] and the
half-duplex performance is lower bounded by the
integral-point-wise linear plot governed by:
\begin{eqnarray}\label{eq:multi antenna ND-SDAFm lower bound}\begin{array}{cccc} d_{\text{ND-SDAFm}}(r) & \geq &
n(m-2r)^2, &  \ \ \ \ 0 \leq r \leq r_\text{coop} \nonumber \\
&=& (m-r)^2, & \ \ \ \ r_\text{coop}<r\leq
m.\end{array}\end{eqnarray} where $r_\text{coop}$ is the
intersection of integral-point-wise linear plots $n(m-2r)^2$ and
$(m-r)^2$. The bound is met by utilizing an $m\times m(n-1)$
horizontally-stacked perfect code (horizontal stacking of $n-1$
independent matrices from an $m\times m$ perfect code) during the
first stage, and a distributed $m(n-1)\times m(n-1)$ perfect code
over the second stage.
\end{prop}
\begin{proof} See Appendix \ref{ssec:appendix multi antennas ND-SDAF
proof}.\end{proof} The proof of approximate universality of the
horizontally-stacked CDA code is found in \cite[Section
V.B]{isit05_explicit}, and is based on the fact that the Hermitian
nature of $XX^\dag$ guarantees that the magnitude of each of the
$m$ ordered eigenvalues of any $XX^\dag$, is lower bounded by the
magnitude of the corresponding eigenvalue of any of the $X^{(k)},
\ \ k=1,2,\cdots,n-1$.

\subsubsection{Approximately universal codes and the most general
relay network setup} We here note that, given any intermediate
relay decoding strategy, it is the case that the generalization of
the above multi-antenna, multi-node, CDA-based scheme, will allow
for an overall network full-duplex D-MG performance, that is
optimal over all other point-to-point encoding schemes, even for
the most general case of having \bit \item an arbitrary number of
receive antennas varying over each terminal\item an arbitrary
number $n_{S}$ and $n_{t_i}$ of transmit antennas varying over the
different nodes $S$ and $\{R_i\}_{i=2}^{n}$  \item arbitrary and
possibly different probability distribution functions for the
fading across the different antenna pairs  \item  an arbitrary
number of users \item an arbitrary decoding strategy  \item an
arbitrary network transmission duration. \eit Specifically, for
some $T$ depending on the network setup, this generalized optimal
network encoding scheme asks for an $n_{S} \times T$ CDA code to
be used for transmission during the first stage, and which will be
part of a $(n_S+\sum_{i=2}^{n}n_{t_i}) \times
(n_S+\sum_{i=2}^{n}n_{t_i})$ CDA code that will be distributed
across all relays, for transmission during the second stage.

\subsection{Proof of D-MG bound on the multi-antenna ND-SDAF
scheme (Proposition \ref{prop:horizontally stacked optimality}
\label{ssec:appendix multi antennas ND-SDAF proof})} The encoding
scheme first asks for the source $S$ to map $m^2(n-1)$ information
elements $\{f_i\}_{i=1}^{m^2(n-1)}$, from a discrete constellation
$\mathcal{A}_m$ of cardinality $|\mathcal{A}_m| =
\mbox{SNR}^\frac{r}{m}$, into $n-1$ codematrices
$\{X^{(k)}\}_{k=1}^{n-1}$, each from an approximately universal
CDA space-time code $\mathcal{X}_m$ of cardinality
$|\mathcal{X}_m| = \mbox{SNR}^{rm}$.  Having done so, $S$
sequentially retransmits $\theta[X^{(1)} \ X^{(2)} \ \cdots \
X^{(n-1)}]$, essentially sending a $m\times m(n-1)$ codematrix $X
= [X^{(1)} \ X^{(2)} \ \cdots \ X^{(n-1)}]$ from the
horizontally-stacked CDA code whose approximate universality was
proved in \cite[Section V.B]{isit05_explicit}). The resulting
guarantee that $P( \ \text{err} \ | \ \text{no outage} \ ) \ \dot
= \ \mbox{SNR}^{-\infty}$, makes it such that, by $t=(n-1)m$, all
intermediate relays in $D(S)$ have \emph{correctly} decoded $X$
and gained full knowledge of $\{f_i\}_{i=1}^{m^2(n-1)}$.
Consequently, each relay in $D(S)$ now re-maps the correct
information into $(m(n-1))^2$ elements from a discrete
constellation $\mathcal{A}_{m(n-1)}$ of cardinality
$|\mathcal{A}_{m(n-1)}| = \mbox{SNR}^\frac{r}{m(n-1)}$, and
eventually re-maps these new information elements into the
codematrix $X_{m(n-1)}$ of an $m(n-1)\times m(n-1)$ CDA code.

In the second stage, the $k^{th}$ transmit antenna
($k=1,\cdots,m$) of intermediate relay $R_j \in D(S)$, transmits
the $((j-1)m+k)^{th}$ row of codematrix $X_{m(n-1)}$.  As a
result, the participating relays essentially construct a
$(|D(S)|+1)m \times mn$ punctured version of the $mn\times mn$ CDA
code.  Due to the residual approximate universality of the CDA
codes, we have that this $|D(S)|m \times mn$ punctured code is
approximately universal over any $|D(S)|m \times n_r$ channel, for
all $n_r$. Furthermore we know that the optimal decoding strategy
will ask for the relays to decode if and only if the corresponding
$m\times m$ channel is not in outage. This, combined with the fact
that the codes, utilized on both stages, have the required
approximate universality and residual D-MG optimality property,
\emph{allows for the proposed encoding scheme to provide for the
optimal full-duplex D-MG performance over the network}.

Narrowing to the specific cases of the Rayleigh fading channel, we
first observe that for $i=0,1,\cdots,n-1$, it is the case that the
expression
\begin{eqnarray*} &\mbox{} &
P(|D(S)|=i)P_\text{out}(im,m) \\ & \dot = &
\bigl(P_\text{out}(m,m)\bigr)^{n-1-i}P_\text{out}(im,m)\\
&\dot = & \mbox{SNR}^{-(m-r)(m-r)(n-1-i)-(im-r)(m-r)}\\ &=&
\mbox{SNR}^{-(m-r)[(m-r)(n-1)-r+ir]}\vspace{-16pt}
\end{eqnarray*}
is maximized for $i=0$.  Consequently, using the dominant summand
approach as in \cite{ZheTse,EliRajPawKumLu}, we conclude that the
optimal full-duplex D-MG performance over the network, is given by
\begin{eqnarray*} P_{\text{err}} & \dot = &
P_\text{out}(m,m)\biggl(\sum_{i=0}^{n-1}
P(|D(S)|=i)P_\text{out}(im,m)\biggr) \\ & \dot = &
P_\text{out}(m,m) P(|D(S)|=0)  \\  & \dot = & P_\text{out}(m,m)
\bigl(P_\text{out}(m,m)\bigr)^{n-1}  \\ & \dot = &
\bigl(P_\text{out}(m,m)\bigr)^n  \\ & \dot = &
\mbox{SNR}^{-n(m-r)^2}.
\end{eqnarray*}
The fact that $2m(n-1)$ time slots are required for the network to
complete the transmission of the $m^2(n-1)$ information symbols,
from a constellation $\mathcal{A}_m$ of cardinality
$|\mathcal{A}_m| = \mbox{SNR}^\frac{r}{m}$, concludes the proof.
\epf

\section{Proofs relating to the ND-RAF scheme
\label{sec:Appendix_proofs for ND-RAF}}

\subsection{The equivalent second stage `two-product' channel \label{ssec:jump to the two-product
channel}} We first note that we can equate having the first linear
dispersion matrix $A_1$ be equal to the identity matrix $I_n$, to
correspond to the direct uncoded transmission between the source
and final destination. Decoding with consideration of the source's
signal, results in savings of one intermediate relay. Having set
$A_1 = I_n$ translates to considering the source as an
intermediate relay, with $g_1 = 1$ known to the receiver and the
transmitter. For uniformity we will assume that the transmitter
does not know $g_1$.  As we proceed, it will become apparent that
this does not affect our analysis. Consequently, we will
henceforth consider $g_i,h_i, \ i=1,\cdots,n$ to be i.i.d,
zero-mean, complex Gaussian random variables.  As in
\cite{JingHass04b}, we now consider the channel equations
(\ref{eq:channelModel1}) and
(\ref{eq:Channel_Matrix_Definition_1}), and observe that matrix
$X$ can be considered as an $n\times n$ codematrix and matrix $H$
as the $1\times n$ channel fading coefficient matrix. The authors
in \cite{JingHass04b,JingHass04c} show that due to the required
unitary nature of the linear-dispersion matrices and due to the
fact that the $h_i$'s are known to the receiver, it is the case
that the effective additive noise elements $W|h_i$ are spatially
and temporally white, zero mean, Gaussian random variables, with
variance
\begin{equation} \label{eq:Var_White_Noise} Var(W|h_i) = (1+\sum_{i=1}^n |h_i|^2)I_n.\end{equation}
In Appendix \ref{ssec:iid nature of fading}, we see that for
unitary $A_i$'s, in the SNR scale of interest, the $W|h_i$ can be
considered, without loss of generality, to be
$\mathbb{C}\mathcal{N}(0,1)$ random variables, conditioned on
considering the $g_i,h_i$ fading coefficients to be i.i.d.
$\mathbb{C}\mathcal{N}(0,\mbox{SNR}^0)$ random variables.  This is
exactly the two-product channel. \epf

\subsection{Proof of the i.i.d nature of fading \label{ssec:iid nature of fading}}
As stated in \cite{JingHass04b,JingHass04c}, ${W|h_i}$ are
spatially and temporally white
${\mathbb{C}\mathcal{N}(0,(1+\sum_{i=1}^n |h_i|^2))}$ random
variables.  The probability density function of ${h = \sum_{i=1}^n
|h_i|^2}$ is $${p(h) = \frac{h^{n-1} e^{-h}}{(n-1)!}}.$$ For
${\epsilon>0}$, $${p(h=\mbox{SNR}^\epsilon) =
\frac{\mbox{SNR}^{(n-1)\epsilon}
e^{-\mbox{SNR}^\epsilon}}{(n-1)!}}$$ and due to the double
exponential term we have that ${p(h=\mbox{SNR}^\epsilon) \dot =
\mbox{SNR}^{-\infty} = 0}$. On the other hand
$$p(h=\mbox{SNR}^{-\epsilon}) = \frac{\mbox{SNR}^{-(n-1)\epsilon}
e^{-\mbox{SNR}^{-\epsilon}}}{(n-1)!},$$ and since
${e^{-\mbox{SNR}^{-\epsilon}}{(n-1)!} \dot \rightarrow 1}$ we have
that $${p(h=\mbox{SNR}^{-\epsilon}) \dot = \mbox{SNR}^{-\epsilon
(n-1)}}$$ which implies that ${Var(W|h_i)\dot = \mbox{SNR}^0}$.

Now we observe that \[\frac{\mathbb{E}[ \|H \theta
X\|_F^2]}{\mathbb{E}[ \|W|h_i\|_F^2]} = n_t \mbox{SNR}=
\frac{n_rn_t Var(g_1h_1)\mathbb{E}[ \|\theta X\|_F^2]}{n_r T
Var(W|h_i)}\] and thus
\[\frac{Var(g_1)Var(h_1)}{Var(W|h_i)}\mathbb{E}[ \|\theta X\|_F^2]
= T \mbox{SNR}.\]  Finally we note that in the SNR scale of
interest, we may interchange the $\mbox{SNR}^0$ and $1$ values and
have $ Var(g_1) = Var(h_1) \dot = \mbox{SNR}^0 $ and $Var(W|h_i) =
1$, since the substitution maintains the power-SNR requirements.

\subsection{Existing performance bounds for the ND-RAF (linear-processing scheme)
(Proposition \ref{prop:expanding old PEP result})
\label{ssec:Appendix proof of expanding old PEP}}
\paragraph{The diversity result from \cite{JingHass04b}}
Having established the whiteness of the additive noise, the
authors in \cite{JingHass04b,JingHass04c} proceed to prove the
following:
\begin{prop}\label{prop:setup For Hassibis Result} (Variant of the main result in
\cite{JingHass04b}) \  \
 Over the effective channel described in
(\ref{eq:channelModel1}), with the channel fading coefficient
matrix as in (\ref{eq:Channel_Matrix_Definition_1}), utilizing
full-rank matrices as in (\ref{eq:Channel_Matrix_Definition_1})
and with white additive noise as in (\ref{eq:Var_White_Noise}), it
is then the case that in the high SNR regime, the pairwise error
probability based diversity is $n$.
\end{prop}
In expanding the above PEP bound to consider multiplexing gains
other than $r=0$, we get:
\begin{prop}\label{prop:expanding old PEP result}
Given random coding, the existing PEP bound in Proposition
\ref{prop:setup For Hassibis Result}, translates to a D-MG bound
(of the entire network) of
\begin{equation*}
d_\text{ND-RAF}(r) \geq (n-1)(1-kr)^{+}+(1-r), \ \ \ \ k
>> 1.
\end{equation*}
Given coding based on approximately-universal codes, the same PEP
bound translates to a D-MG bound for the entire network of
\begin{equation*}d_\text{ND-RAF}(r) \geq (n-1)\left(1-\frac{4n-1}{n-1}\right)^{+}+(1-r).\end{equation*}
\end{prop}
\begin{proof}
It is not difficult to see that in the high-SNR regime, the PEP
diversity of the second stage of the network scheme, as given in
\cite{JingHass04b,JingHass04c}, is a function only of SNR and of
the minimum eigenvalue of any difference of any two codematrices
(at a given rate), and is upper bounded as:
\begin{equation}\label{eq:Hassibi_PEP_HighSNR}PEP \dot \leq \mbox{SNR}^{-n}
(l'_n)^{-n}\end{equation} where $l'_n$ is the smallest eigenvalue
of the power normalized version of the code where the codematrices
$X'$ are of the form ${X' = \frac{X}{E}}$ such that
${\mathbb{E}[\|X'\|_F^2] \dot = 1}$.

For the channel model given as ${Y = \theta HX+W}$, with
${\mathbb{E}[\|X\|_F^2] \dot = E^2}$ and $\theta$ such that
${\mathbb{E}[\|\theta X\|_F^2] \dot = \mbox{SNR}}$, it was shown
in \cite{EliRajPawKumLu} that a non-vanishing determinant
${\det(\Delta X \Delta X^\dag) \dot \geq \mbox{SNR}^0}$,
guarantees a minimum partial eigenvalue product of
$${\prod_{i=n-j}^n l_i \geq (E^2)^{-(n-j-1)}}$$ and minimum
code-eigenvalue of ${l_n \ \dot \geq \  (E^2)^{-(n-1)}}$.

Interestingly, in the normalized case of
$\mathbb{E}(\|X^{'}\|_F^2) = 1$ with ${l'_i = \frac{l_i}{E^2}}$,
the minimum partial eigenvalue product becomes
\[{\prod_{i=n-j}^n l'_i \geq (E^2)^{-n} \ \ \ \  \forall j}\]
which accentuates nicely the fact that code performance depends
solely on its smallest eigenvalue. For when the ${n\times n}$ CDA
code $\mathcal{X}$ maps $n^2$ information symbols from a discrete
constellation $\mathcal{A}$, we have that
\[{E^2 = |\mathcal{A}| = |\mathcal{X}|^{\frac{1}{n^2}} =
(2^{Rn})^{\frac{1}{n^2}} = (2^{n r\log
\mbox{SNR}})^{\frac{1}{n^2}} = \mbox{SNR}^{\frac{r}{n}}}.\] As a
result ${l'_n \dot \geq (\mbox{SNR}^{\frac{r}{n}})^{-n} =
\mbox{SNR}^{-r}}$ which brings us to
\[PEP\dot \leq \mbox{SNR}^{-n}\mbox{SNR}^{-r}\]  The overall probability of error
is then \[{P_e\leq |\mathcal{X}| (PEP) =
\mbox{SNR}^{nr}\mbox{SNR}^{-n}\mbox{SNR}^{-r}}\] which allows for
lower-bounding the D-MG tradeoff of the two-product channel as
\begin{equation}\label{eq:PEPbased_DMG}d_{\text{equiv}}(r)\geq n-2rn\end{equation} implying a maximum diversity
${d_{\text{equiv}}(0) = n}$ and maximum multiplexing gain of
${r_{\text{equiv}} \geq \frac{1}{2}}$. Consequently, a temporally
disjoint first and second stage result in cooperation applying at
the solution of $n(1-4r) = 1-r$.  This concludes the proof.
\end{proof}
Given approximately universal codes, the PEP based D-MG bound can
only guarantee that cooperation will apply for
$r<\frac{n-1}{4n-1}$ which, given some rate $R$, translates to
having cooperation only after $\mbox{SNR}>2^{\frac{4n-1}{n-1}R}$.
Given finite duration random codes, cooperation might require
$\mbox{SNR}>2^{kR}, \ \ \ \ k>>1$.

\subsection{Lemma on the error contribution region of perfect
codes \label{ssec:Appendix lemma on error contr region of perfect
codes}}
\begin{lem}\label{lem:Error_Contribution_Region} For $ \lambda_n  = HH^\dag=\sum_{i=1}^n
\|h_i\|^2\|g_i\|^2 := \mbox{SNR}^{-\mu} $ being the only non-zero
eigenvalue of the two-product channel, then the corresponding
outage region which at high SNR equals the error contribution
region of the perfect codes, is given by:
\begin{equation} \label{eq:region_Of_Error_Contribution}
\mathcal{B} = \{ \mu \geq 1-r\}.
\end{equation}
\end{lem}
\begin{proof}
Directly from the mismatched eigenvalue theorem presented in
\cite{EliRajPawKumLu}, we know that the minimum Euclidean distance
between any two codematrices after the action of the channel, is
lower bounded as
\begin{equation}\label{eq:Euclidean_lmin_lambdaMax} d^2_E = \|\theta H \Delta X
\|_F^2 \  \dot \geq  \ \theta^2 \lambda_n l_n\end{equation} where
$l_n$ corresponds to the smallest code eigenvalue of ${\Delta X
\Delta X^\dag}$.  For the case of the perfect code $\mathcal{X}$,
which maps $n^2$ information symbols from a set $\mathcal{A}$, we
know that $\theta^2 = \mbox{SNR}^{1-\frac{r}{n}}$, $l_n \dot \geq
\mbox{SNR}^{-\frac{r}{n}(n-1)}$. As a result,
\begin{equation} \label{eq:PEP_mismatched_2}
 d_E^2(\mu) \geq
\mbox{SNR}^{1-\frac{r}{n}}\mbox{SNR}^{-\frac{r}{n}(n-1)}
\mbox{SNR}^{-\mu}:= \mbox{SNR}^c.\end{equation} Observing that the
pairwise error probability, given a channel, is upper bounded as
\[PEP(\mu) \leq e^{-d_E^2(\mu)}\] allows for the entire
probability of error, given a channel $\mu$, to be upper bounded
as:
\[P(err|\mu)\dot \leq |\mathcal{X}| \cdot PEP(\mu) = \mbox{SNR}^{nr} e^{-\mbox{SNR}^c}.\]
It is clear that \[ \lim \limits_{\mbox{SNR}\rightarrow \infty}
\mbox{SNR}^{nr}e^{-\mbox{SNR}^c} = \mbox{SNR}^{-\infty} = 0, \
\forall c>0, \] simply because we have a double exponentially
decreasing term multiplied with a polynomially increasing term. As
a result, we have the error contribution region, being defined as
the set $\mathcal{B}$ of eigenvalues $\mu$ such that
$P(err|\mu)\neq 0$. This region requires that $c\leq 0$.
Consequently, (\ref{eq:PEP_mismatched_2}) gives that ${c = 1-r-\mu
\leq 0}$ which concludes the proof.
\end{proof}

\subsection{Lemma on the effective pdf in the two-product channel \label{ssec:Appendix lemma on effective pdf of two-product channel}}
\begin{lem} \label{lem:Final_Channel_PDF}
For $h_i,g_i, \ i=1,2,\cdots,n$ being
$\mathbb{C}\mathcal{N}(0,\mbox{SNR}^0)$ random variables, then the
probability density function of $\lambda_n = \sum_{i=1}^n
\|h_i\|^2 \|g_i\|^2$ is upper bounded as
\begin{eqnarray}\label{eq:finalPDF}
f_\lambda(\lambda) \dot \leq \lambda^{n-1}
\end{eqnarray}
and for ${\lambda_n = \mbox{SNR}^{-\mu}}$
\begin{equation} \label{eq:finalSimplifiedPDF}f_{\mu}(\mu) \dot \leq \mbox{SNR}^{-\mu n}.\end{equation}
\end{lem}
\begin{proof}
We first recall that for $x,y$ being two independent random
variables and for $z=xy$, we have that ${f_{z,x}(z,x) =
\frac{1}{|w|} f_{x,y}(x,\frac{z}{x})}$ and thus
\[
 f_z(z) = \int\limits_{-\infty}^{\infty} \frac{1}{|w|}
f_{x,y}(x,\frac{z}{x}) dx\] Now consider $h_i,g_i$ being two
independent identically distributed ${\mathbb{C}\mathcal{N}(0,k)},
\ k\dot = \mbox{SNR}^0$, random variables and thus ${x =
\|h_i\|^2}$, ${y = \|g_i\|^2}$, two i.i.d. exponential random
variables with ${f_x(x) = e^{-kx}, \ f_y(y) = e^{-ky}}$,
${f_{x,y}(x,y) =e^{-kx}e^{-ky}}$ and ${f_{x,y}(x,\frac{z}{x}) =
e^{-kx}e^{-k\frac{z}{x}}}$, ${x,y,z\in \mathbb{R}^+}$.  As a
result,
\[{f_z(z) = \int \limits_{0}^{\infty} \frac{1}{w} e^{-kw}
e^{-k\frac{z}{w}} dw= \int \limits_{0}^{\infty} \frac{1}{w}
e^{-k(w+\frac{z}{w})}}dw.\]  From equation 3.471.9 of
\cite{GR_Integral_Tables} we see that ${f_z(z) =
2K_0(2k\sqrt{z})}$ i.e.
\[f_{\|h_i\|^2\|g_i\|^2}(z) = 2K_0(2k\sqrt{z})\] where ${K_0(\cdot)}$
corresponds to the modified Bessel function of the second kind.

From \cite[Section 6.6]{nrC}, we observe that
\[ \lim \limits_{x\rightarrow 0} \frac{K_0(x)}{-log(x)}\rightarrow
1.\] Observing that for ${x  \tilde{\geq} 0.5}$, ${K_0(x)
\tilde{\leq} \frac{\pi}{\sqrt{2 \pi x}} e^{-x}}<1$ is always
finite (and decreasing in a double-exponential rate), allows one
to interchange $\mbox{SNR}^0$ and $1$.  Using the absolute unity,
we also note that as ${\epsilon \rightarrow 0}$ then
\[\frac{K_0(1-\epsilon)}{-log(1-\epsilon)}\approx \epsilon^{-1}\]
We then choose a guaranteed to exist $\epsilon$ such that
${\frac{K_0(1-\epsilon)}{-log(1-\epsilon)} = k \dot =
\mbox{SNR}^0}$. It is easy to see that $1-\epsilon$ is closer to
$1$ than to $0$. This combined with the monotonically decreasing
nature of $K_0(x)$, gives that
\[{\int \limits_{0}^{1-\epsilon} K_0(x) dx \dot > \int
\limits_{1-\epsilon}^1 K_0(x)dx }\] and as a result of the
dominant summand effect
\[\int \limits_{0}^1 K_0(x) dx \dot = \int \limits_{0}^{1-\epsilon} K_0(x)
dx +0.\]  Consequently, the term ${\int_{1-\epsilon}^1 K_0(x) dx}$
 can be substituted by the smaller term $\int_{1-\epsilon}^1 -
\log(x) dx$.  Combined with the fact that ${K_0(1)  \dot =
K_0(\mbox{SNR}^0) \dot = \mbox{SNR}^0}$ and that
${\frac{K_0(x_1)}{-\log(x_1)}<\frac{K_0(x_2)}{-\log(x_2)} \
\forall x_1<x_2}$, it is the case that for all ${z\dot \leq 1}$,
we get \[ \int \limits_{0}^z K_0(\sqrt{z})dz  \dot \leq  -k \int
\limits_{0}^z \log(\sqrt{z})dz \dot = - \int \limits_{0}^z
\log(\sqrt{z})dz \]

Recall that if $ \lambda \in \mathcal{B}$ then $\lambda\dot \leq
\mbox{SNR}^0$.  The fact that all the summands of the channel
eigenvalue ${\lambda = \sum_{i=1}^n |g_i|^2|h_i|^2}$ are positive,
suggests that if $\lambda \in \mathcal{B}$ then $|g_i|^2|h_i|^2
\dot \leq \mbox{SNR}^0$.  This justifies limiting our attention to
the $f_{|g_i|^2|h_i|^2}(x), \ x\leq \mbox{SNR}^0$.  As a result,
for $h_i,g_i$ being ${\mathbb{C}\mathcal{N}(0,\mbox{SNR}^0)}$
random variables, then the cumulative distribution function of
${\|h_i\|^2\|g_i\|^2}$, up to $\mbox{SNR}^0$, is upper bounded as
\begin{equation} \label{eq:CDF_of_single_product1}
P(\|h_i\|^2\|g_i\|^2\leq z) \dot \leq z(1-\log(z))\end{equation}
since
\begin{eqnarray*}
F_z(z) & \dot = & - \int\limits_{0}^z \log(\sqrt{t})dt \dot =
 -\frac{1}{2} \int\limits_{0}^z \log(t)dt \nonumber \\
&=& \frac{1}{2}(t-t \log(t))|_0^z = z-z \log(z)+0\log(0)
\end{eqnarray*}
and since ${0\log(0) = 0}$, we get that
\begin{equation}\label{eq:CDF of one dimensional product} P(\|h_i\|^2\|g_i\|^2<z) \dot = z(1-\log(z)).\end{equation}
To complete the proof of the main Lemma
\ref{lem:Final_Channel_PDF}, we use the following proposition:
\begin{prop}\label{prop:Hypercube proposition}
Consider $n$ independent identically distributed random variables
${a_1,\cdots,a_n}$ with ${a_i>0, \ \forall i}$. It is then the
case that for ${\lambda =\sum_{i=1}^n a_i}$ then
${P(\lambda<t)\dot \leq [P(a_i)<t]^n}$. \end{prop} \begin{proof}
Sketch of proof of Proposition \ref{prop:Hypercube proposition}:
In the $n=2$ case, the ${(a_1,a_2)}$ pairs such that ${a_1+a_2<t}$
are the points inside the triangle formed by the positive
$a_1$-axis, the positive $a_2$-axis and the line ${a_2=t-a_1 =
f(a_1)}$. As a result, for a certain $a'_2$, it is the case that
${a'_2 + a_1<t, \ \forall a_1<f^{-1}(a'_2)}$. The corresponding
triangle is completely enclosed in the ${t\times t}$ square whose
lower-left corner is at the ${(0,0)}$ point. As a result
${P(a_1+a_2<t)\leq P(a_1<t)P(a_2<t) = (P(a_1<t))^2}$. For the
$3$-dimensional case, the triangle is replaced by a pyramid which
is enclosed in a $3$-dimensional cube with edge of length $t$, and
in the general $n$-dimensional case the $n$-dimensional pyramid is
enclosed in an $n$-dimensional hypercube again with edge of length
$t$.  This concludes the proof of Proposition \ref{prop:Hypercube
proposition}.\end{proof} Given the above proposition we continue
from (\ref{eq:CDF of one dimensional product}) and can now deduce
that, for the cumulative distribution function
${P(\|h_i\|^2\|g_i\|^2\leq z) \ \dot \leq \ z(1-\log(z))}$
presented in equation (\ref{eq:CDF_of_single_product1}) and for
${\lambda = \sum_{i=1}^n \|h_i\|^2\|g_i\|^2 }$, we have that
\[P(\lambda \leq z)\leq [P(\|h_i\|^2\|g_i\|^2<z)]^n  \ \dot \leq \ [z(1-log(z))]^n\]
Consequently \[F_\lambda(\lambda) \ \dot \leq \
[\lambda(1-log(\lambda))]^n\] Keeping in mind that the range of
interest is ${\lambda \leq \mbox{SNR}^0}$ and that
${F_\lambda(\mbox{SNR}^0)\dot = \mbox{SNR}^0}$, then for ${\lambda
= \mbox{SNR}^\delta}$, ${\mbox{SNR} \rightarrow \infty, \
1>\delta>0}$, it is the case that ${(\mbox{SNR}^\delta
(1-\log(\mbox{SNR}^\delta)))^m  \ \dot =  \ (\mbox{SNR}^\delta)^m,
\ m\geq 0}$. As a result, the cumulative distribution function
(cdf) is expressed as a polynomial with only one term as
\[{F_\lambda(\lambda)\dot \leq \lambda^n}\] which allows us, after differentiation, to
apply the existing upper bound inequality of the cdf to the pdf,
and get that
\begin{eqnarray}\label{eq:finalPDF}
f_\lambda(\lambda)  \ \dot \leq  \ \lambda^{n-1}
\end{eqnarray}
For ${\lambda = \mbox{SNR}^{-\mu}}$, ${\mu = \frac{-\log
(\lambda)}{\log(\mbox{SNR})}}$ then ${ f_\mu(\mu) =
\frac{f_\lambda(\lambda)}{\frac{d\mu }{d\lambda}}}$ with
${\frac{d\mu}{d\lambda} =
-\frac{1}{\log(\mbox{SNR})}\frac{1}{\lambda}}$. As a result,
\begin{eqnarray}
f_\mu(\mu)  \ \dot \leq  \ \mbox{SNR}^{-\mu n}\log(\mbox{SNR})
\end{eqnarray}
As in \cite{ZheTse}, the fact that
\[\lim_{\mbox{SNR}\rightarrow \infty} \frac{\log((\log(\mbox{SNR})))}{\log (\mbox{SNR})}
\rightarrow 0\] indicates that the $\log(\mbox{SNR})$ term does
not contribute to the SNR exponent of the probability of error,
and as a result
\begin{equation} \label{eq:finalSimplifiedPDF}f_{\mu}(\mu)  \ \dot \leq \  \mbox{SNR}^{-\mu n}\end{equation}
which concludes the proof of Lemma \ref{lem:Final_Channel_PDF}.
\end{proof}

\subsection{Proof of the antenna limitations of the relay network \label{ssec:Antenna limitations of network}}
The fact that the relays can only perform linear processing,
prohibits multiple antennas at the source since one cannot
linearly process matrices in a meaningful way due to the additive
nature of the received signal at the intermediate relays.
Furthermore the intermediate relays can only have one
receive-transmit antenna due to lack of source-to-relay channel
information.

\subsection{Relation of number of mapped symbols to $ r_{max} $
\label{ssec:Proof of relation of number of mapped symbols to
rmax}}
\begin{prop}\label{prop:mn_discrete_symbols_rmax_is m}
Consider a full-diversity  $n\times T$, $T\geq n$, space-time code
$\mathcal{X}$ that maps, through linear combining, on the average
$mT$ information elements from a discrete constellation
$\mathcal{A}$. Given that the code operates in the Rayleigh fading
with $n$-transmit and $n_r$ receive antennas, with $n_r\geq m$, or
operates in the two-product channel with $n$-transmit antennas ($m
\leq 1$), then such a code can achieve a maximum multiplexing gain
of $ r_{max} \leq m $.
\end{prop}

\begin{proof}
We have that
 $|\mathcal{X}| = 2^{RT} =  2^{rT\log_2\mbox{SNR}} =
\mbox{SNR}^{rT} = |\mathcal{A}|^m$ which implies that
$|\mathcal{A}| \dot = \mbox{SNR}^\frac{r}{m}$ and since the
constellation is discrete $\mathbb{E}[\|\alpha\in
\mathcal{A}\|^2]\dot = |\mathcal{A}|$. The fact that each element
$X_{i,j}$ of a code matrix is a linear combination, with
coefficients independent of SNR, of elements of $\mathcal{A}$,
gives us that $\mathbb{E}[\| X_{i,j} \|^2] = |\mathcal{A}| =
\mbox{SNR}^\frac{r}{m}$.  For $\theta$ such that
$\mathbb{E}[\|\theta H X\|_F^2] = \mathbb{E}[\|\theta X\|_F^2]
\dot = \mbox{SNR}$ we have that $\theta^2 =
\mbox{SNR}^{1-\frac{r}{m}}$.

Without loss of generality we can assume that there exist two
codeword matrices $X_1, X_2\in \mathcal{X}$, with $X_i$ mapping
the information $nm$-tuple $\{\alpha_i,0,0,\cdots,0\}$, where
$\alpha_i \dot = \mbox{SNR}^0$.  As a result, the determinant and
trace of the difference matrix $\Delta X$, is a polynomial of
degree less than $n$ over $\alpha = a_1-a_2 \dot = \mbox{SNR}^0$,
with coefficients independent of SNR, i.e. $$\det(\Delta X \Delta
X^\dag ) \ \dot = \  Tr(\Delta X \Delta X^\dag)  \ \dot = \
\mbox{SNR}^0$$ and thus with all its eigenvalues $l_i \ \doteq \
\mbox{SNR}^0$. The corresponding pairwise error probability $PEP(X
\rightarrow X')$, in the Rayleigh fading channel, then serves as a
lower bound to the codeword error probability $P_e$, i.e.,
\begin{eqnarray*} P_e &\geq & PEP(X \rightarrow X^{'}) \ \doteq \
\frac{1}{\prod_{j=1}^n [1 + \frac{\theta^2}{4} l_j ]^{n_r}} \\
& \doteq &\mbox{SNR}^{-n_rn (1- \frac{r}{m})} \end{eqnarray*}
\begin{eqnarray}
\Rightarrow \ d(r) \ \leq n_rn \left(1- \frac{r}{m} \right)
\end{eqnarray}
which proves the claim for the Rayleigh fading case.  Remark
\ref{note:two product no better than Rayleigh} concludes the
proof. \end{proof}

\subsection{Relation between rate reduction due to half-duplex and number of mapped symbols
\label{ssec:Rate_Reduction_Due_To_Duplex}}
\begin{lem}\label{lem:Rate_Drop_Due_To_Duplex}
Consider a distributed $n\times n$ space-time code, with one of
its linear-dispersion matrices equal to the identity matrix, and
with D-MG performance over the two-product channel given by
$d_{\text{eq}}(r)$. Furthermore assume that the space-time code
carries $mn$ information elements from a discrete constellation.
It is then the case that the D-MG tradeoff of the entire
half-duplex constrained linear-processing relay network utilizing
this code, is given by
\begin{equation} \label{eq:Half_Duplex_Drop_In_r}
d_{\text{network}}(r) = d_{\text{eq}}(r(m+1)). \end{equation}
\end{lem}
\begin{proof}
We first observe that linear-processing does not allow for the
intermediate relays to extract information from received signals
since the distributed code $\mathcal{X}$ can only hold linear
combinations of the received symbols. This forces the source to
transmit the $mT$ information symbols one at a time, and the
intermediate relays to transmit over $T$ time slots, without
information extraction and re-encoding. This means that the lack
of decoding at the intermediate relays does not translate to
arbitrarily high rates. It is in fact the case that the rates are
limited both by the capacity-outage corresponding to the
dimensions of the second stage channel, as well as from the fact
that the transmission during the second stage will have to take
place over $\frac{mT}{m} = T$ time slots.

For $r$ corresponding to the multiplexing gain of the second
stage, it is the case that $|\mathcal{X}| = \mbox{SNR}^{rT}$ and
as a result the information constellation satisfies
\[|\mathcal{A}|\dot \leq \mbox{SNR}^{\frac{r}{m}}.\]

Consequently, in the $T=n$ case, for $d(r)$ corresponding to the
transmission over the $n$ time slots of the second-stage, it is
then the case that $d(r(m+1))$ corresponds to the entire $mn+n$
time-slots (first and second stage duration). \end{proof}

\section{Proofs relating to the D-RAF scheme
\label{sec:Appendix_proofs for D-RAF}} From \cite{ElGamalDUplex},
we see that due to the independence of the information symbols,
the power constraints are satisfied when
$$b_i \leq \sqrt{\frac{\mbox{SNR}}{|g_i|^2\mbox{SNR}+1}}.$$ We denote by
$\phi_i$ and $\xi_i$ the exponential orders of $h_i$ and $g_i$
respectively, i.e.
\begin{equation}\label{eq:exponential orders of D-RAF fades}
h_i \dot = \mbox{SNR}^{-\phi_i}, \ \ \ g_i \dot =
\mbox{SNR}^{-\xi_i}\end{equation} and as in \cite{ElGamalDUplex},
we observe that
\begin{equation}\label{eq:imposibility of high fading}p_{\phi_i}(\phi_{i}) = \biggl\{
\begin{array}{lr} \mbox{SNR}^{-\infty}, & \phi_{i}<0 \\ \mbox{SNR}^{-\phi_i}, & \phi_i \geq 0
\end{array}.\end{equation}  The same holds for $\xi_i$.  As a result, in
accordance with \cite{ZheTse,EliRajPawKumLu,ElGamalDUplex} and
without loss of generality we will limit our attention to
$\xi_i,\phi_i\in \mathbb{R}^{+}$. Given that the probability that
$\xi_i,\phi_i\in \mathbb{R}^{-}$ can be considered to be
arbitrarily small, we may restrict the amplification factors to be
$b_i = 1$.

\subsection{Proof of Proposition \ref{prop:multiple inter
relay main result of Azarian D-RAF}\label{ssec:appendix proof of
n=2 D-AAF Azarian}} Beginning with the single intermediate relay
case, a frame (\ref{eq:amp and for set of equations1}), is given
by
\begin{eqnarray*}
\underline{y} = \! [ y_1 \   y_2 ]\! & = &
\underline{s}+\underline{n}
\\ & = & [h_1 x_1 \ \ h_2 g_2x_1+h_1x_2] +[w_1 \ \
h_2v_{2,1}+w_2] \nonumber.\end{eqnarray*} For compactness we will
denote SNR by $\rho$.  For \begin{equation} \label{eq:equivalent
channel G2} G_2 =
\begin{array}{|cc|} h_1 & 0\\ h_2g_2 & h_1
\end{array}\end{equation} denoting the equivalent channel defined
by the above equations, we have $\Sigma_s = \rho G_2G_2^\dag $
denoting the covariance matrix of the signal and fading part of
the observed vector at the destination, with the expectation taken
over the signaling set and the fading coefficients held fixed.
Furthermore, $\Sigma_n  \  \dot =  \ \begin{array}{|cc|} 1 & 0\\
0 & 1+ |h_2|^2 \end{array}$ denotes the covariance matrix of the
additive noise part of the observed vector at the destination.
Using the fact that the largest eigenvalue $l_{\max}$ of
$\mathbb{E}[x_ix^*_j]$ satisfies $l_{\max} \dot = \rho$ and
directly from (\ref{eq:exponential orders of D-RAF
fades})-(\ref{eq:imposibility of high fading}), the mutual
information during a single $2$-length frame is given in
\cite{ElGamalDUplex} to be
\begin{eqnarray}\label{eq:mutual information 0}I(\underline{x};\underline{y}|h_1,h_2,g_2) & \dot
= & \log_2 \det(I_2+\rho GG^\dag\Sigma_n^{-1}) \\  & \dot = &
\max\{2(1-\phi_1)^+,1-\phi_2-\xi_2\} \log_2\rho.\nonumber
\end{eqnarray}
We now consider the outage event, that is the event where the
corresponding maximum mutual information allowed by a set of
fading coefficients is less than the rate of transmission $R =
r\log_2\rho$.  That is
\begin{equation*}
\mathcal{O} = \bigl\{ (h_1,h_2,g_2)\in \mathbb{C}^3 \ | \
I(\underline{x};\underline{y}|h_1,h_2,g_2) <R T_f \bigr\}
\end{equation*}
where $T_f$ is the duration of the frame.  Furthermore, due to
(\ref{eq:imposibility of high fading}), we are only interested in
\begin{multline}\label{eq:outageExpression amplify and forward1}
\mathcal{O}^+   = \\
  =  \biggl\{ (\phi_1,\phi_2,\xi_2)\in \mathbb{R}^{+3}  | \
\max\{2(1-\phi_1)^+,1-\phi_2-\xi_2\}\bigr\}
 <2r \biggr\}.
\end{multline}
The probability of outage is given by \begin{eqnarray}
P_\text{out}(r) & \dot = & \rho^{-d_\text{out}(r)} =
\int\limits_{(\phi_1,\phi_2,\xi_2)\in \mathcal{O}}
P(\phi_1,\phi_2,\xi_2) d\phi_1 d\phi_2 d\xi_2  \nonumber \\
& \dot=& \sup\limits_{(\phi_1,\phi_2,\xi_2)\in \mathcal{O}^+}
\rho^{-(\phi_1+\phi_2+\xi_2)}.
\end{eqnarray}
The last equation comes from the dominant term approach in
\cite{Hardy_Orders_Of_Infinity}, extensively used in
\cite{ZheTse}\cite{ElGamalDUplex}\cite{EliRajPawKumLu}, or
directly from Varadhan's lemma \cite{DemZei}. Consequently
\begin{equation}\label{eq:dout(r) in half duplex amp and forward}
d_\text{out}(r) =  \inf\limits_{
\max\{2(1-\phi_1)^+,1-\phi_2-\xi_2\} <2r } \{  \phi_1+\phi_2+\xi_2
\}
 \end{equation}
which readily leads to the final expression
$$d_\text{D-RAF}(r) = (1-r)+(1-2r)^+.$$

Now moving to the multiple intermediate relay case, we again
assume Gaussian random coding across the frame, and introduce
$\Sigma_{s_i}$, $i=2,3,\cdots,n$
\begin{eqnarray*}
\Sigma_{s_i}= \rho G_iG_i^\dag  &=&
\rho \ \begin{array}{|cc|} |h_1|^2 & h_1h_{i+1}^{*}g_{i+1}^{*} \\
h_1^{*}h_{i+1}g_{i+1} & |h_1|^2+|h_{i+1}|^2|g_{i+1}|^2
\end{array}
\end{eqnarray*}
and $\Sigma_{n_i}$, $i=2,3,\cdots,n$
\begin{eqnarray*}
\Sigma_{n_i} & = &
\begin{array}{|cc|} \sigma_w^2 & 0\\ 0 & \sigma_w^2+
|h_{i+1}|^2\sigma_v^2 \end{array}  \  \dot = \  \begin{array}{|cc|} 1 & 0\\
0 & 1+ |h_{i+1}|^2 \end{array}.
\end{eqnarray*}
As a result, the block diagonal covariance matrices are now
\begin{eqnarray} \Sigma_s =\begin{array}{|ccc|}
\Sigma_{s_2} & 0_{2\times 2} & \cdots \\
& \ddots & \\
 0_{2\times 2} & \cdots & \Sigma_{s_{n}}
\end{array},  \Sigma_n =\begin{array}{|ccc|}
\Sigma_{n_2} & 0_{2\times 2} & \cdots \\
& \ddots & \\
 0_{2\times 2} & \cdots & \Sigma_{n_{n}}
\end{array}
\end{eqnarray}
Proceeding as in the single intermediate relay case, concludes the
proof for the probability of outage corresponding to the
equivalent channel
\begin{eqnarray}
\label{eq:equivalent channel multiple intermediate relays} G
=\begin{array}{|cccc|}
G_2 & 0_{2\times 2} & \cdots & 0_{2\times 2}\\
0_{2\times 2} & G_3 & \cdots & 0_{2\times 2}\\
& & \ddots & \\
 0_{2\times 2} & 0_{2\times 2} & \cdots & G_{n}
\end{array}.
\end{eqnarray}

\epf

\subsection{Proof of Theorem \ref{thm:D AAF all relays}
\label{ssec:appendix proof n=2 vectorized perfect D-RAF}} For the
single intermediate relay case, we use the equivalent
representation
\begin{equation}\label{eq:equivalent representation}Y = HX+W\end{equation} of the scheme i.e.
\begin{multline} \left[ \begin{array}{cc} y_1 & y_3 \\
y_2 & y_4 \end{array} \right] = \left[ \begin{array}{cc} h_1 & 0 \\
g_2 h_2 & h_1 \end{array} \right] \left[ \begin{array}{cc} x_1 &
x_3 \\
x_2 & x_4 \end{array} \right] \\  +  \ \left[ \begin{array}{cc}
w_1 &
w_3 \\
h_2v_{2,1}+w_2 & h_2v_{2,3}+w_4  \end{array} \right]
\end{multline} and modify to
reflect noise whitening by \begin{eqnarray} \Sigma & = & \mathbb{E} \left\{ \left[ \begin{array}{c} w_1 \\
h_2 v_{2,1}+w_2
\end{array} \right] \left[ \begin{array}{cc} w_1^{*} &
h_2^{*}v_{2,1}^{*}+w_2^{*} \end{array} \right] \right\} \nonumber \\ &= & \left[ \begin{array}{cc} 1 & 0 \\
0 & |h_2|^2+1 \end{array} \right]\end{eqnarray}
to the equivalent whitened model \begin{multline} \Sigma^{-1} \left[ \begin{array}{cc} y_1 & y_3 \\
y_2 & y_4 \end{array} \right] \\ =  \Sigma^{-1} \left[ \begin{array}{cc} h_1 & 0 \\
g_2 h_2 & h_1 \end{array} \right] \left[ \begin{array}{cc} x_1 &
x_3 \\
x_2 & x_4 \end{array} \right] \  \\+  \ \Sigma^{-1} \left[
\begin{array}{cc} w_1 &
w_3 \\
h_2v_{2,1}+w_2 & h_2v_{2,3}+w_4  \end{array}
\right].\end{multline}

For $$ \Sigma^{-1} \left[
\begin{array}{cc} w_1 &
w_3 \\
h_2v_{2,1}+w_2 & h_2v_{2,3}+w_4  \end{array} \right]  :=  \left[
\begin{array}{cc} u_1 & u_3 \\
u_2 & u_4 \end{array} \right]$$ and $$ \ \ \Sigma^{-1}
\left[ \begin{array}{cc} h_1 & 0 \\
g_2 h_2 & h_1 \end{array} \right]  =  \left[ \begin{array}{cc} h_1 & 0 \\
\frac{g_2h_2}{1+|h_2|^2} & \frac{h_1}{1+|h_2|^2}
\end{array} \right] $$ we also let
\begin{eqnarray*} \left[ \begin{array}{cc} z_1 & z_3 \\
z_2 & z_4 \end{array} \right] & = & \Sigma^{-1} \left[ \begin{array}{cc} y_1 & y_3 \\
y_2 & y_4 \end{array} \right] \\ & = &  \left[ \begin{array}{cc} h_1 & 0 \\
A g_2 h_2 & h_1 A \end{array} \right] \left[
\begin{array}{cc} x_1 &
x_3 \\
x_2 & x_4 \end{array} \right] \ \\ & & + \ \left[ \begin{array}{cc} u_1 & u_3 \\
u_2 & u_4 \end{array} \right] \end{eqnarray*} where $ A \ = \
\frac{1}{1+|h_2|^2} $.  We now define $$
H_{\text{eff}} \ = \ \left[ \begin{array}{cc} h_1 & 0 \\
A g_2 h_2 & h_1 A \end{array} \right] $$ and proceed to find the
outage probability by calculating
\begin{eqnarray*} & \mbox{}&
\det (I + \rho H_{\text{eff}}H_{\text{eff}}^{\dagger} ) =  \\ & =
& \det \left[
\begin{array}{cc} 1+ \rho | h_1|^2 & \rho h_1 (A g_2 h_2)^{*}
\\
\rho (A h_2 g_2)h_1^{*} & 1 + \rho \{ |A g_2 h_2 |^2 + | h_1 A |^2
\}
\end{array} \right] \\
& = &  \biggl(1+\rho |h_1|^2\biggr)\biggl(1+ \rho \{ |A g_2 h_2
|^2 + | h_1 A |^2 \}\biggr)  \\ &-&  \rho^2 \left(
|h_1|^2 |Ag_2g_2|^2 \right) \\
& = & 1 + \rho |h_1|^2 + A^2 \biggl( \rho  |h_2g_2|^2 + \rho |
h_1|^2 + \rho^2 | h_1|^4 \biggr). \end{eqnarray*} This gives that
\begin{multline*} P_{\text{out}}(r) \\ =  Pr \biggl( \log \bigl[ 1
+ \rho |h_1|^2  + A^2 \bigl(\rho |h_2g_2|^2 + \rho | h_1|^2 +
\rho^2 | h_1|^4 \bigr) \bigr]\\ < 2 r \log (\rho)
\biggr)\end{multline*} and then for
$$ \overline{h}_1 := |h_1|^2 , \ \ \overline{h}_2 :=
|h_2|^2 \ \ \overline{g}_2 := |g_2|^2,$$ the probability of outage
is given by
$$ \int_{\{\overline{h}_1,\overline{h}_2,\overline{g}_2\} \in {\cal O}} e^{ - [\overline{h}_1+\overline{h}_2+\overline{g}_2]} d\overline{h}_1
d\overline{h}_2 d\overline{g}_2$$ where
\begin{multline*}
{\{\overline{h}_1,\overline{h}_2,\overline{g}_2\} \in {\cal O}}
\Leftrightarrow\\\Leftrightarrow \log \bigl[ 1+ \rho
\overline{h}_1 + A^2 \bigl( \rho \overline{h}_2\overline{g}_2+
\rho \overline{h}_1+ \rho^2 \overline{h}_1^2 \bigr) \bigr] \ \leq
\ 2 r \log (\rho )
\end{multline*}
i.e., \begin{multline*}
{\{\overline{h}_1,\overline{h}_2,\overline{g}_2\} \in {\cal O}}
\Leftrightarrow \\ \Leftrightarrow 1+ \rho \overline{h}_1 + A^2
\bigl( \rho \overline{h}_2\overline{g}_2+ \rho \overline{h}_1+
\rho^2 \overline{h}_1^2 \bigr)
 \
\leq \ \rho^{2r}.\end{multline*} At this point, we can see that
since $$ \overline{h}_1, \overline{h}_2, \overline{g}_2 \ \dot{
\leq } \ \rho^{0}$$ (from \cite{ElGamalDUplex}), we can
approximate outage as $$ {
\{\overline{h}_1,\overline{h}_2,\overline{g}_2\} \in {\cal O}}
\Rightarrow 1+ \rho \overline{h}_1 + \rho
\overline{h}_2\overline{g}_2+\rho \overline{h}_1+\rho^2
\overline{h}_1^2  \ \leq \ \rho^{2r}
$$ and can thus write
\[
P_\text{out} =
\int_{\{\overline{h}_1,\overline{h}_2,\overline{g}_2\} \in {\cal
O}} e^{ - [\overline{h}_1+\overline{h}_2+\overline{g}_2]}
d\overline{h}_1 d\overline{h}_2 d\overline{g}_2.
\]
It follows that the effective (whitened) channel $H_{\text{eff}}$
has the same outage as does the original channel $H$, and this
outage curve is met by the use of the approximately universal
$2\times 2$ CDA code. For the general case of having $n-1$
intermediate relays, the whitened equivalent channel described by
(\ref{eq:amp and for set of equations1}) is exactly given by
(\ref{eq:equivalent channel multiple intermediate relays}).  This
channel's outage curve was found in \cite{ElGamalDUplex}, shown
here in Proposition \ref{prop:multiple inter relay main result of
Azarian D-RAF}, and is exactly met by the corresponding
approximately universal $2(n-1)\times 2(n-1)$ CDA code. \epf

\section{Distributed perfect space-time codes
\label{sec:AppendixPerfectCodesDistirbutabilityAndApproximateUniversality}}
\subsection{Distributed perfect space-time codes\label{sec:Appendix_Perfect_Codes}}
As shown in \cite{EliRajPawKumLu}, the basic elements of a CDA
space-time code are the number fields $\mathbb{F}, \mathbb{L}$,
with $\mathbb{L}$ a finite, cyclic Galois extension of
$\mathbb{F}$ of degree $n$. For $\sigma$ being the generator of
the Galois group $\mbox{Gal}(\mathbb{L}/\mathbb{F})$, we let $z$
be some symbol that satisfies the relations $ \ell z \ = \ z
\sigma(\ell) \ \ \ \forall  \ \  \ell \in \mathbb{L} \ \ \  \mbox{
and } \ \ \ z^n=\gamma $ for some `non-norm' element $\gamma \in
\mathbb{F}^*$ such that the smallest integer $t$ for which
$\gamma^t$ is the relative norm $N_{\mathbb{L}/\mathbb{F}}(u)$ of
some element $u$ in $\mathbb{L}^*$, is $n$. From there we
construct a cyclic division algebra $D= \mathbb{L} \oplus z
\mathbb{L} \oplus \hdots \oplus z^{n-1}\mathbb{L}$.  A space-time
code $\mathcal{X}$ can be associated to $D$ by selecting the set
of matrices corresponding to the left-regular representation of
elements of a finite subset of $D$. For an arbitrary choice of
integral basis $\{\beta_i\}_{i=0}^{n-1}$ for a submodule in the
ring of integers $O_\mathbb{L}$ of $\mathbb{L}$ over $\mathbb{F}$,
the elements of the signaling set are of the form ${\ell_i =
\sum_{j=0}^{n-1} f_{i,j} \beta_j, \ \ \in \ell_i\in \mathbb{L}, \
f_{i,j}\in O_\mathbb{F}}$.  As a result, the CDA code-matrix form
is
\begin{equation} \label{eq:distributed_CDA} X= \sum_{j=0}^{n-1}
\Gamma^j \biggl( diag \bigl( \underline{f}_{j}\cdot G \bigr)
\biggr)  = \sum_{u=1}^n \underline{f}A_u
\end{equation}
with ${ \underline{f}_j =[ f_{j,0} \  f_{j,1} \ \cdots \ f_{j,n-1}
]}$ and the concatenated $\underline{f}$ now being the
$n^2$-length sequence of $O_\mathbb{F}$ information elements
leaving the transmitter node (after normalization). Each
$n^2\times n$ matrix $A_u$ (corresponding to relay $u$) is created
by first letting
\\$A_{n,i} = diag \left[
\begin{array}{ccccc} \beta_i & \sigma(\beta_i) & \sigma^2(\beta_i)
&\cdots & \sigma^{n-1}(\beta_i)
\end{array}\right]_{n\times n}$ for $i=0,1,\cdots,n-1$ and then
recursively creating $A_{u,i} = \Gamma^{n-u} A_{n,i} , \  \
u=1,2,\cdots,n$ where
 \fontsize{11}{13}
 \begin{eqnarray}\label{eq:G_Matrix}G & = & \begin{array}{|ccc|}
\sigma^0(\beta_0)  &  \cdots &   \sigma^{n-1}(\beta_{0})\\
&  \vdots & \\
\sigma^{0}(\beta_{n-1}) & \cdots &
 \sigma^{n-1}(\beta_{n-1})\\
\end{array} \\[5pt] \label{eq:GammaMatrix}  \ \  \Gamma & = & \begin{array}{|cccc|}
0 & 0 & \cdots  & \gamma\\
1 & 0 & \cdots  & 0\\
0 & 1 & \cdots  & 0 \\
 & \vdots & & \end{array} \  \ \  A_u =
\begin{array}{|c|} A_{u,0}\\\vdots\\
A_{u,n-1}\end{array} \ \ \\[5pt]  X &=&
\begin{array}{|c|}
\underline{f}A_1\\
\vdots\\
\underline{f}A_n
 \end{array}
\end{eqnarray}
\normalsize

The perfect code requirement that the lattice generator matrix $G$
and the power sharing matrices $\Gamma^j, \ j=0,1,\cdots,n-1$ in
(\ref{eq:GammaMatrix}), be unitary matrices guarantees that $
A_u^\dag A_u = I$, essential for the code's information
losslessness.

\begin{center}
ACKNOWLEDGEMENT
\end{center}
The authors would like to thank Solomon Golomb, Robert Guralnick
and Giuseppe Caire for their useful comments.

\bibliographystyle{IEEEbib}

\end{document}